\documentclass[amsmath,prl,aps,twocolumn]{revtex4}
\usepackage{amsthm,amsfonts,graphicx,verbatim}
\usepackage{hyperref}

\newcommand{\Tr}{{\rm Tr}}

\newcommand{\D}{{\rm d}}

\newcommand{\be}{\begin{equation}}
\newcommand{\ee}{\end{equation}}
\newcommand{\bea}{\begin{eqnarray}}
\newcommand{\eea}{\end{eqnarray}}

\newcommand{\la}{\langle}
\newcommand{\ra}{\rangle}

\renewcommand{\epsilon}{\varepsilon}
\def\nn{\nonumber\\}

\usepackage{color}
\definecolor{Red}{rgb}{0.8,0,0}
\definecolor{Black}{rgb}{0,0,0}

\newcommand{\new}{\begingroup\color{Red}}
\newcommand{\old}{\endgroup}

\begin{document}
\title{Radiation matter entanglement}
\author{I. Klich}%$^{1}$ A. Retzker$^{2}$}%
%\email{klich@caltech.edu}
\affiliation{Department of Physics, University of Virginia, Charlottesville, VA 22904}
%\\ (2) Department of Physics.... }
\begin{abstract}
The quantization of the electromagnetic field in the presence of material bodies, at zero temperature is considered. It is shown that a dielectric does not act as thermal bath for the field and yields a non-trivial non-thermal mixed state of the field. The properties of this state and its entropy are studied. The dependence of the second Renyi entropy of the field on the distance between dispersive objects is shown to decay as $R^{-4}$ for generic bodies. \end{abstract}
\maketitle

\section{Introduction}
The study of quantum correlations and entanglement in many body quantum systems has become a major theme in recent years. In particular, many-body systems are studied in terms of the reduced density matrices describing the state of constituents of the system. This point of view gives rise to various questions such as the question of thermalization: When can a part of a quantum system be considered as effectively in a thermal state? 

Naturally, thermal density matrices appear in systems weakly coupled to a thermal heat bath. Moreover, it has recently been argued that a canonical thermal state may arise for typical pure states of the system+bath Hilbert space after tracing out the bath \cite{goldstein2006canonical,popescu2006entanglement}. However, many physical systems of interest are not typical and violate some of the assumptions leading to thermalization. In particular, the system considered here, that of a boson radiation field interacting with a dielectric medium, characterized by a dielectric function $\epsilon(\omega)$, has been shown \cite{klich2012prlentanglement} to be described by a density matrix which is not thermal for the typical frequency dependence of the dielectric function $\epsilon(\omega)$.

The dielectric function serves as a convenient way to encode the particular long wavelength properties of the material \cite{LP}. The dielectric susceptibility of the material, and it's relations with various response functions serves to define an effective action for the field, and through it enables one to describe the actual state of it. 

%In particular, the thermalizion hypothesis which states that the long time behavior of out-of equilibrium states may be described by thermal averages has attracted much recent interest. 
%In this paper we concentrate on important physical system, namely that of the electromagnetic field in contact with dielectric bodies. 
In this paper we continue the investigation, initiated in \cite{klich2012prlentanglement} of the reduced density matrix of a boson field in contact with a dispersive medium. Such a field may describe many other situations, such as phonons in a solid. The resulting photonic density matrix, of a field interacting with realistic materials, is in general not in a pure state and should be described in terms of a mixed state density matrix. This situation is in stark contrast to idealized Dirichlet or Neumann boundaries, which are consistent with a field Hamiltonian. The main observation is that the resulting mixed state of the field is not thermal, which may be a consequence of the nature of the model we consider, which is, in essence, integrable, and thus not typical.  
%We do this by describing the effective density matrix of the state and it's associated entropy.

As a measure of the field-matter entanglement, we use the entanglement entropy of the field. The entropy of radiation coupled to matter has been considered in numerous works. However, usually, the focus is on a single degree of freedom coupled to a bath of oscillators: For example, the entropy of a spin in the spin-boson model within the frame work of the Caldeira-Legget model \cite{caldeira1981influence} was considered in \cite{hur2008entanglement} and the entanglement of a single radiation mode with an array of spins was studied in \cite{lambert2004entanglement} for the Dicke model . The entanglement between spatially separated intervals of vacuum (or ground state of a spin chains) has been considered in \cite{marcovitch2009critical,calabrese2009entanglement}. Here we have a system distinct from these works, and more akin to situations studied in  macroscopic electrodynamic effects such as the Casimir and Van der Waals interactions between dielectric bodies. 

Upon considering the entanglement entropy of the radiation field in contact with a material, it becomes clear that the entropy suffers from a UV divergence, and is thus explicitly cutoff dependent. Instead, in order to look for universal long-range features, we consider the distance dependent part of the second Renyi entropy of a field interacting with two distinct bodies. 
The second Renyi entanglement entropy is substantially more tractable analytically the full Von-Neumann entropy, but often contains the correct scaling behavior. Recently, Renyi entanglement entropies have been computed for numerous systems. For example, the ability to numerically access ${\cal S}_2$ has been used to probe entropy and topological entropy in recent works \cite{hastings2010measuring, song2011entanglementHeisenberg,ju2012entanglement}. 

In particular, we estimate the second Renyi entropy and find that:
\bea&\label{Renyi decay}
{\cal S}_2(A\cup B)-{\cal S}_2(A)-{\cal S}_2(B)=\nn & -\omega _{{pA}}^2\omega _{{pB}}^2V_AV_B\frac{2\pi ^4}{\omega_{0}^2R^4}+O({1\over R^{6}}),
\eea
where 
$
{\cal S}_2(A)=-\log\Tr \rho_{\phi,A}^{2 }.$ Here $\rho_{\phi,A}$ is the density matrix of a field in contact with a body $A$, after the body degrees of freedom have been integrated out. This result is of interest, as it shows that the decay of entanglement in this model is very slow compared to typical power laws in quantum electrodynamics such as interaction energies between dielectric bodies in the Casimir-Polder regime and compared to the toy-model studied in  \cite{klich2012prlentanglement}.

The paper is organized as follows. After a brief introduction of the problem, we proceed to review the description of density matrices of Gaussian states and their entropy. We then use this formalism to describe the state of a field in material medium using the equal time correlations of the field and field momenta operators. In the following section, we turn to the description of the distance dependent part of the entropy of a field interacting with two objects. We derive a formal expression for this entropy. Finally, we study the distance dependence of the second Renyi entropy obtaining the result \eqref{Renyi decay}.
 
\section{Gaussian effective action}
%\section{Action of the electromagnetic field}
The basic principles for describing field fluctuations in the vicinity of heated bodies have been comprehensively explored since the early days of electrodynamics  (see e.g.  \cite{LP,levin1967theory}). Within this approach, the macroscopic  field interaction with the material is described through the dielectric response function $\epsilon({\omega},{\bf x})$. Following this logic, here we study a simplified scalar field version of the electromagnetic field action %\eqref{actionA}
%We study a simplified scalar model described by
%As a simplified model, consider the action: 
\begin{eqnarray}\label{action}
S={1\over 4\pi}\int \D^3x{\D\omega}\phi_{\omega}^*({\bf x})[{\omega}^2\epsilon({\omega},{\bf x})-\nabla^2]\phi_{\omega}({\bf x}).
\end{eqnarray}
Throughout the paper we will also use the susceptibility $\chi$, which is related to the dielectric function $\epsilon$ by $\epsilon({\omega},{\bf x})=1+\chi(\omega,{\bf x})$.

When the permittivity is independent of $\omega$, the action is local in time, and one can easily quantize the associated scalar action
{\it assuming} the conjugate momentum $\pi_{\phi}$ can be expressed in terms of $\dot{\phi}$ and doesn't depend on external fields.
Such an action follows from the %a (Hermitian) field
Hamiltonian $H={1\over 4\pi}\int\D^3x [{\pi^{2}\over \epsilon({\bf x})}+(\nabla\phi)^{2}]$. It describes, at zero temperature, a pure state, and as such will have no entropy. Note that $x$ dependence of $\epsilon$ does not interfere with this property: it just means that the field has a spatially non-uniform mass, but can still be describe in terms of a Hamiltonian.

The situation is fundamentally different if $\epsilon$ is $\omega$ dependent. The non-locality of the action \eqref{action} in time signals that our system is coupled to external degrees of freedom which have been integrated out, yielding a non trivial temporal response kernel. In such a case, the system cannot be in a pure state even at zero temperature, implying that our radiation field is entangled with the matter fields. 

The effective action \eqref{action} can be obtained from integrating out other fields. To keep in mind a simple model for such a procedure we have in mind a bosonic  field $\phi$ coupled with a matter field $\psi$. We consider the following typical action:
\bea\nonumber&\label{Action with matter}
S_{\text{pure}}=\frac{1}{2}\int _{-\infty }^{\infty }{dt }\Big\{\int d^3x[\phi\left(-\partial_{t}^2+\nabla ^2\right)\phi]+\nn &\int_{B} d^3x[\frac{1}{8\pi ^2}\psi\left(-\partial_{t}^2-\omega _0{}^2\right)\psi _{\omega }+\omega_p  \psi(\partial_{t}\phi)]\Big\}\eea
% This action may be viewed as a {\it purification} of the state $\rho_{\phi}$ of $\phi$,  i.e. we write our state as $\rho_{\phi}=\Tr_{\psi}|\Omega\ra\la\Omega|$ for some pure state $|\Omega\ra$ in the larger $\phi,\psi$ space.
The action \eqref{Action with matter} corresponds to the form $\chi (\omega )=\frac{\omega _p{}^2}{\omega _0{}^2-\omega ^2}$ of the response of $\phi$ to a transparent, but dispersive medium.

Dissipation may be introduced in a similar way, but requires coupling to an infinite bath of oscillators for each field degree of freedom, as done, e.g. in the Caldeira-Legget model \cite{caldeira1981influence}. 

When quantizing a system starting from an effective action such as \eqref{action} it is important to keep in mind the following subtle point: the quantization procedure cannot be complete without additional information on the system. In our case, we will need the conjugate momentum operators to the field and their correlations. It is impossible to obtain those from the effective action alone.  The reason is that while  the action gives us full information about the field correlators, it does not define uniquely the momentum correlations: The momentum correlations are extracted from the time dependence of the field correlations through equations of motion, but these depend on the particular way the field is coupled to the matter.

To illustrate this point, we show that there may be some ambiguity on how to correctly choose those. For example, the terms $\phi \partial _t \psi$ and $
-\psi \partial _t \phi$ in a field Lagrangian, while classically the same, as they are related by a full time derivatives, yield the same effective action for $\phi$ upon integrating $\psi$ out but, are not quantized in the same way. Indeed, let us look at the following simple example. Consider the two Lagraniangs:
\bea
L_1=\frac{1}{2}\left(\dot{\phi}^2-\omega _0{}^2 \phi ^2\right)+\frac{1}{2}\dot{\psi}^2-B \psi ^2+a\dot{\psi} \phi 
\eea
\text{and }
\bea
L_2=\frac{1}{2}\left(\dot{\phi}^2-\omega _0{}^2 \phi ^2\right)+\frac{1}{2}\dot{\psi}^2-B \psi ^2-a \dot{\phi} \psi 
\eea
which differ by a total derivative: $L_1-L_2=a\frac{d}{\text{dt}}({\phi \psi})$.  Let us check, that when canonically quantizing them we get different Hamiltonians:
From the Lagrangian $L_{1}$, { we get the canonical: }
$P_ \phi =\dot{\phi}${ and }
$P_ \psi =\dot{\psi}+a \phi $ {While }$L_2$ gives us: 
$P_ \phi=\dot{ \phi}-a$
{  and }$P_ \psi =\dot{\psi}$.

Clearly, the time dependent correlation functions $\la\phi(t)\phi(t')\ra$ computable from the action \eqref{action}, will yield different results for $\la\pi_{\phi}(t)\pi_{\phi}(t')\ra$ when computed from the equation of motion obtained from of the full Lagrangians $L_{1}$ and $L_{2}$.

In this paper we proceed choosing $\pi_{\phi}=\dot{\phi}$  motivated by the usual lagrangian coupling the electromagnetic field $A$ with material fields through $A\cdot J$.

\section{Gaussian states and their entropy}
For a general many-body density matrix, determination of the state requires, in principle, the knowledge of all matrix elements of the density matrix. These scale exponentially with the number of degrees of freedom and for interacting systems become intractable very quickly. However, fortunately, for states described by Gaussian field theories the situation is considerably simpler. 
Indeed, by virtue of Wick's theorem, all correlation functions, and thus all matrix elements can be obtained from the two point functions of the field. Thus, our task is to use the two point functions in order to represent the general state of the field.
To proceed, in this section we review the method of calculating entropies of Gaussian states (see, e.g. \cite{botero2003modewise,plenio2007introduction}) from correlation functions. The material is standard, but for convenience, we choose review it here in detail.

Consider a scalar field with $n$ degrees of freedom $\phi_{n}$ and conjugate momenta $\pi_{n}=-i\partial_{\phi_{n}}$. It is convenient to bunch these together, defining the vector:
\bea(O_1,..O_{2n})=(\phi_1,..\phi_n,\pi_1,..\pi_n).\eea
Using this vector, the canonical commutation relations $[\phi_{n},\pi_{m}]=i\delta_{nm}$ may be  expressed as:
\bea\label{commutation relations}
[O_j,O_k]=i \sigma_{j,k}
\eea
where $\sigma$ is the $2n\times2n$ block matrix 
\bea\sigma= \left(
\begin{array}{cc}
0 &  I_{n}  \\
-I_{n}  &  0
\end{array}
\right)\eea
and we set $\hbar=1$ for convenience.

The commutation relations \eqref{commutation relations} as well as the hermiticity of the canonical field and momenta operators are preserved under the group of symplectic transformations $Sp(2n)$, the set of real matrices $W$ such that $W \sigma W^{T}= \sigma$.

As mentioned above, we would like to use the two point functions of the field to determine it's state. To do this, let us first define the the covariance matrix:
\begin{eqnarray}\label{Covariance Mat}
\gamma_{jk}=2Re\la(O_{j}-\la O_{j}\ra)(O_{k}-\la O_{k}\ra)\ra.
\end{eqnarray}
The operators $O_{i}$ can always be redefined so that $\la O_{j}\ra=0$. In the rest of the paper we will assume our Gaussian field and field momenta have been correctly defined to have this property.

%To find the canonical, normal mode representation associated with this state, 
Next, we bring 
 $\gamma$ into  the canonical Williamson normal form:
 \bea\label{cannonical form of gamma}
 W \gamma W^{T}=\left(
\begin{array}{cc}
{\tilde{\mu}}  & 0 \\
 0 & {\tilde{\mu}}
\end{array}
\right)\equiv M
 \eea
Where ${\tilde{\mu}}$ is a diagonal matrix ${\tilde{\mu}}=diag(\mu_{1},...\mu_{n})$ and $W\in Sp(2n)$ . The $\mu_{i}$ are called the ``symplectic eigenvalues'' of $\gamma$, and they are equal to the positive eigenvalues of $i\sigma \gamma$.

To see this last property, let us check that if $i\sigma\gamma\psi=\mu\psi$ then $\mu^{2}$ is an eigenvalue of $M^{2}$. To see this one can explicitly check that $\sigma M\sigma M=-M^{2}$, therefore:
\begin{eqnarray*} &  M^{2}=-\sigma M\sigma M=-\sigma W \gamma W^{T} \sigma W \gamma W^{T} =\nn & -W^{-T}W^{T}\sigma W \gamma  \sigma  \gamma W^{T} =-W^{-T} \sigma   \gamma  \sigma  \gamma W^{T}.
\end{eqnarray*}
We conclude that \bea M^{2}(W^{-T}\psi)=  W^{-T}  (i\sigma   \gamma )^{2}W^{T} W^{-T}\psi=\mu^{2}(W^{-T}\psi).\eea

%Alternatively, if we are {\bf assured} that the expectation value of $\la x_{i}p_{j}\ra$ is imaginary, than we just have to compute the eigenvalues of $iGH$, where $G$ is the $\phi\phi$ correlator, and $H$ is the $\pi\pi$ correlator, then take square roots of the eigenvalues.

An additional, technical simplification, occurs, if we assume that no $\la\phi\pi\ra$ correlations are present, i.e. $\la\phi\pi\ra=0$. In this case the symplectic eigenvalues are equal to square roots of eigenvalues of ${\cal G}{\cal H}$, where ${\cal G}$ and ${\cal H}$ are field and field momentum two point functions, respectively. To see this property, consider a $\gamma$ which is block diagonal of the form:
\bea \gamma =2\left(
\begin{array}{cc}
 {\cal G} & 0 \\
 0 & {\cal H}
\end{array}
\right)\eea
Assume that $i\sigma \gamma\psi=\mu\psi$. It follows that $\gamma^{2}\psi=\mu^{2}\psi$. Computing, explicitly, \bea
(i \sigma \gamma )^2 =4\left(
\begin{array}{cc}
  {\cal H} {\cal G} & 0 \\
 0 &  {\cal G} {\cal H}
\end{array}
\right)
\eea
Thus
\bea
(i \sigma \gamma )^2 \left(
\begin{array}{c}
 \psi_1 \\
 \psi_2
\end{array}
\right)=4\left(
\begin{array}{c}
    {\cal H} {\cal G} \psi_1 \\
 {\cal G} {\cal H} \psi_2
\end{array}
\right)=4\mu^2\left(
\begin{array}{c}
 \psi_1 \\
 \psi_2
\end{array}
\right)\eea
and we conclude that $\mu^{2}$ is an eigenvalue of  $ {\cal G} {\cal H}$ (and, equivalently, of $ {\cal H}{\cal G}$).

Next, we find the Gaussian state associated with the covariance matrix $\gamma$ by comparing to the covariance matrix of a general quadratic Hamiltonian in a thermal state. 

Consider a quadratic Hamiltonian for coordinate and momenta $\phi,\pi$
\bea H=\psi ^TM \psi~~~~~~;~~~~~\psi =\left(
\begin{array}{c}
 \phi \\
 \pi
\end{array}
\right),
\eea
{where M is symmetric and real. We find the normal form }
\bea
S^TM S=\left(
\begin{array}{cc}
{\Lambda}  & 0 \\
 0 & {\Lambda}
\end{array}
\right)\eea
{ and let: }
$
\psi =S\tilde{\psi }$.
Then \bea
H=\tilde{\psi }^T\left(
\begin{array}{cc}
{\Lambda}  & 0 \\
 0 & {\Lambda}
\end{array}
\right)\tilde{\psi }
\eea
In terms of the components $\tilde{\phi}_i,\tilde{\pi}_i $ of $\tilde{\psi}$, the Hamiltonian breaks into independent oscillators of the form:
\bea&
H=\Sigma _iH_i\text{     };\text{     }\nn & H_i=\Lambda_i\left(\tilde{\phi}_i{}^2+\tilde{\pi}_i{}^2\right)=2 \Lambda _i\left(a_i^+a_i+\frac{1}{2}\right)
\eea
{where }$
a_i^+=\frac{1}{\sqrt{2}}\left(\tilde{\phi}_i-i \tilde{\pi}_i \right)$ are canonical bosonic creation operators.

Let us compute the finite temperature expectation values of  
$\tilde{\phi}_i,\tilde{\pi}_i $. At an inverse finite temperature $\beta $ we have:
\begin{eqnarray}&
\left\langle \tilde{\phi}_i^2\right\rangle =\frac{1}{2}\left\langle \left(\tilde{a}_i{}^++\tilde{a}_i\right){}^2\right\rangle =\nn & \frac{1}{2}\left\langle 2\tilde{a}_i{}^+\tilde{a}_i+1\right\rangle =\tilde{n}_i+\frac{1}{2}=\frac{1}{e^{2\beta  \Lambda_i} -1}+\frac{1}{2}
\eea
where $\tilde{n}_i=\tilde{a}_i{}^+\tilde{a}_i$ is the number occupation of mode $i$.
Thus: 
\begin{eqnarray*}
\tilde{\gamma }=2\left(
\begin{array}{cc}
 \left\langle \tilde{\phi}_i\tilde{\phi}_j\right\rangle  & \left\langle \tilde{\phi}_i\tilde{\pi}_j\right\rangle  \\
 \left\langle \tilde{\pi}_i\tilde{\phi}_j\right\rangle  & \left\langle \tilde{\pi}_i\tilde{\pi}_j\right\rangle 
\end{array}
\right)=2\left(
\begin{array}{cc}
 \frac{1}{e^{2\beta  {\Lambda}  } -1}+\frac{1}{2} & 0 \\
 0 & \frac{1}{e^{2\beta  {\Lambda}  } -1}+\frac{1}{2}
\end{array}
\right)
\end{eqnarray*}
Finally, we rotate to the original basis, so that 
\bea \label{cannonical form of thermal} &
\gamma =2\left(
\begin{array}{cc}
 \left\langle \phi_i \phi_j\right\rangle  & \left\langle \phi_i \pi_j\right\rangle  \\
 \left\langle \pi_i \phi_j\right\rangle  & \left\langle \pi_i \pi_j\right\rangle 
\end{array}
\right)=\nn & 2\left(S^T\right)^{-1}\left(
\begin{array}{cc}
 \frac{1}{e^{2\beta  {{\Lambda}}  } -1}+\frac{1}{2} & 0 \\
 0 & \frac{1}{e^{2\beta  {\Lambda}  } -1}+\frac{1}{2}
\end{array}
\right)S^{-1}.
\eea
Comparing \eqref{cannonical form of gamma} with  \eqref{cannonical form of thermal}, and setting $\beta=1$ we see that:
$${\mu_i\over 2}=\frac{1}{e^{2 \Lambda_{i} } -1}+\frac{1}{2}.$$ Note that the choice of $\beta$ is arbitrary, since one can always rescale simultaneously $\beta$ and $\Lambda$ while keeping their product constant. Inverting the relation we find 
$
e^{2 \Lambda _{\text{eff}} } -1=2(\mu_i-1)^{-1}
$, and finally:
\bea
2 \Lambda_i=\log\frac{\mu_i+1}{\mu_i-1}.\eea
% ({\bf ADD SOMEWHERE  what is the importance of $\gamma$ it being real?})

Thus, we conclude that given a covariance matrix $\gamma$, and assuming this is a Gaussian state, then we can describe the state of the system as:
\bea 
\label{Effective rho normal}
\rho_{}=Z^{-1}e^{-\sum_{i} 2\Lambda_{i}{a^{+}}_{i}a_{i}}
~;~2\Lambda_{i}=\log{\mu_{i}+1\over \mu_{i}-1}.
\eea

Next, let us compute the entropy of this state in terms of the symplectic eigenvalues $\mu_{i}$. Since the density matrix factors into independent modes, we may consider first a single $i$ mode. For a non interacting bosonic state defined as 
$
\rho _i=\frac{1}{Z_i}e^{-2\Lambda_{i}  a^+{}_ia_i }$, 
the eigenvalues are labeled by the number of bosons occupying the mode. The occupation probabilities are:
\bea
p_{i,n}=\frac{1}{Z_i}\xi_{i}^{n},
\eea
where  for $n=0,..\infty $ and we defined $\xi_{i}=e^{-2\Lambda_{i}}$. {Let us compute the normalization:}
\bea
Z_i=Tr e^{-2\Lambda_{i} a^+{}_ia_i }= \sum _{n=0}^{\infty } \xi_{i}^{n}=\frac{1}{1-\xi_{i}}
\eea
The entropy is:
\bea&
S_i=-\sum_{n=0}^{\infty} p_{i,n}\log p_{i,n}= -\sum _{n=0}^{\infty }  \frac{\xi_{i}^{n}}{Z_i}\log \left( \frac{\xi_{i}^{n}}{Z_i}\right) =\nn & \log  Z_i-\frac{\log  \xi_{i}}{Z_i}\sum _{n=0}^{\infty }  n \xi_{i}^{n}=\log  Z_i-\frac{ \xi_{i} \log   \xi_{i}}{Z_i} \partial_{\xi_{i}} Z_i(\xi_{i})=\nn & -\log (1-\xi_{i})-\frac{ \log  \xi_{i}}{(1-\xi_{i})}
\eea
Substituting $\xi_{i}=e^{-2\Lambda_{i}}={\mu_{i}+1\over\mu_{i}-1}$ we find 
that the entropy of the mode is given by:
\bea
S_i={\mu_{i} +1\over 2}\log{\mu_{i} +1\over 2}-{\mu_{i}-1\over 2}\log{\mu_{i}-1\over 2}
\eea
Since the modes are independent, the total entropy is simply:
\begin{eqnarray}&\label{Entropy formula}
{\cal S}=\sum_{i} h(\mu_{i}) ~~~;~~~ \nn & h(\mu)={\mu +1\over 2}\log{\mu +1\over 2}-{\mu-1\over 2}\log{\mu-1\over 2}
\end{eqnarray} 

Let us check how this works for the simple harmonic oscillator in the ground state of a Hamiltonian $H={1\over 2m}p^{2}+{\omega^{2}m\over 2}x^{2}$. We know that the state is pure and so should have zero entropy.
Here we have a single degree of freedom so $\gamma$ is a $2\times 2$ matrix, which is easily computed:
\begin{eqnarray}
\gamma=2\begin{pmatrix}
    \la xx\ra  &   0 \\
  0    &  \la pp\ra\end{pmatrix}=\begin{pmatrix}
   {1\over m\omega}  &   0 \\
  0    &  { m\omega}
\end{pmatrix}
\end{eqnarray}
now we have:
\begin{eqnarray}
i\sigma\gamma=i\begin{pmatrix}
 0    &   { m\omega} \\
  -{1\over m\omega}    &  0
\end{pmatrix}
\end{eqnarray}
with eigenvalues $\pm 1$. As remarked before, the symplectic eigenvalues of $\gamma$ are the positive eigenvalues of $i\sigma \gamma$. Therefore the entropy is given by:
\begin{eqnarray}
S=h(1)=0
\end{eqnarray}

More generally, we can compute the Renyi entropies, defined as:
\bea \label{Renyi entropy def} S_{\alpha }=\frac{1}{1-\alpha }\log\Tr \rho^{\alpha }.\eea
where, in particular, 
$S_1=-\Tr \rho \log  \rho $
is the Von Neumann entropy. 

First computing, as before, the entropy per mode, we have that:
 \bea
 \Tr \rho_{i}^{\alpha}={1\over Z_{i}^{\alpha}}Tr e^{-2\alpha \Lambda_{i} a^+{}_ia_i }
 \eea
 And thus:
 \bea&
 S_{\alpha,i}=\frac{1}{1-\alpha }\log{(1-\xi_{i})^{\alpha }\over 1-\xi_{i}^{\alpha}}
 \eea
Expressing $S_{\alpha,i}$ explicitly in terms of the symplectic eigenvalues we find:
\bea\label{Renyi entropy using symplectic eigenvalues}
S_{\alpha,i }=\frac{\log \left(\left(\frac{\mu_{i} +1}{2}\right)^{\alpha }-\left( \frac{\mu_{i} -1}{2}\right)^{\alpha }\right)}{-1+\alpha },\eea
and the final expression for the Renyi entropy $S_{\alpha }$, is obtained by summing over all the modes: 
\bea&
S_{\alpha }=\sum_{i} h_{\alpha}(\mu_{i}) ~~~;~~~ \nn & h_{\alpha}(\mu)=\frac{\log \left(\left(\frac{\mu_{i} +1}{2}\right)^{\alpha }-\left( \frac{\mu_{i} -1}{2}\right)^{\alpha }\right)}{-1+\alpha }.\eea
In particular, we will be interested in the second Renyi entropy, given explicitly as:
\bea\label{Renyi2 explic}&
S_{\alpha }=\sum_{i} \log(\mu_{i}).\eea

\section{Density matrix of the field in a homogenous medium}
In the next sections, we explore the state of a scalar field using the formalism elucidated above for Gaussian states. To do so we compute the elements of the covariance matrix and it's symplectic eigenvalues.

We note that carrying the symplectic transformations needed to diagonalize \eqref{Covariance Mat} implied by our action \eqref{action}, in general may be a hard task. However, the important case of a homogenous system allows us to proceed analytically since for a translationally invariant medium the symplectic eigenvalues can be labeled by momentum.

In the case of a homogenous medium, the action \eqref{action} becomes:
\begin{eqnarray}\label{homogenous action}
S={1\over 4\pi}\int {\D^3k\over (2\pi)^{3}} {\D\omega}\phi_{\omega}^*({\bf k})[{\omega}^2\epsilon({\omega},{\bf k})+{\bf k}^{2}]\phi_{\omega}({\bf k}).
\end{eqnarray}

As explained above, the density matrix of the field can thus be written as:
\bea\label{rho effective homeogenous}
\rho_{field}=Z^{-1}e^{-\sum_{k} 2\Lambda_{k}a^{+}_{k}a_{k}}
\eea
where $a^{+}_{k}$ is a bosonic creation operator labeled by momentum $k$. 

For the models we consider here, the system doesn't have $\la\phi\pi\ra$ correlations, thus we proceed by computing the field two point functions ${\cal G}$ and field momentum correlations ${\cal H}$.
As explained above, the desired symplectic eigenvalues $\mu_{k}$ are square roots of the eigenvalues of the matrix ${\cal G}{\cal H}$. In a homogenous space, both ${\cal G}$ and ${\cal H}$ are diagonal in momentum space, and we are left with the task of computing the momentum space correlators.

We thus have:

Explicitly, to compute the correlation functions, we will use that for a single harmonic oscillator, governed by an action
\bea
\int _{-\infty }^{\infty }\frac{\text{d}\omega}{2\pi }A(\omega ) \phi(\omega )\phi^*(\omega ),
\eea
 we have:
 \bea
\langle \phi(t)\phi(0)\rangle =\int _0^{\infty }\frac{\hbar  e^{-i \omega  t}}{A(i\omega )}\frac{\text{d}\omega}{2\pi }
\eea
Therefore, taking account that the momentum is a good quantum number we can write:
$\mu_{k}=2\pi^{-1}(g_{{\bf k}}h_{\bf k})^{{1/2}}$, where
 the action \eqref{action} allows us to compute, per $k$ mode:
\begin{eqnarray}\label{corrphi}
g_{k}=\la \phi_{\bf k} \phi_{-\bf k}\ra=
\int_0^{\infty} \D\omega 
 {1\over  {\omega^2   \epsilon({\bf k},i\omega)+k^2}}.
\end{eqnarray}
Similarly, using time point splitting and the equation of motion $\pi=\dot{\phi}$
 \begin{eqnarray}\label{corrpi}&
h_{k}=\la \pi_{\phi}^2\ra_{\bf k}=\lim_{t'\to t}\partial_{t}\partial_{t'}
\int_0^{\infty} \D\omega 
 {e^{i\omega(t'-t)}\over  {\omega^2   \epsilon({\bf k},i\omega)+k^2}}=\nn &
\int_0^{\infty} \D\omega 
 {1\over  {\omega^2    \epsilon({\bf k},i\omega)+k^2}}(k^2+\chi(i|\omega|)).
\end{eqnarray}

\begin{comment} Noting that \bea&
\left\langle \phi _k(t)\phi _{-k}(0)\right\rangle = \int d^3x d^3x'\langle \phi (x,t)\phi (x',0)\rangle e^{i k (x-x')}=\nn & \int d^3x_+ d^3x_-\left\langle \phi \left(x_-,t\right)\phi (0,0)\right\rangle e^{i k x_-}=L^d\int  d^3x_-{\cal G}(x_-,0)e^{i k x_-},\eea
we finally find that the eigenvalues of the operators  ${\cal G}{\cal H}$ are of the form:
\bea
L^{-2d}g_{k}h_{k}
\eea
where $L^{d}$ is the volume of space. We then have:\end{comment}
As usual, quantizing the $k$s according to $k={2\pi\over L}(n_{1},..n_{d})$ where $n_{i}$ are integers, and $L\rightarrow \infty$ is the linear size of space, we have that
\begin{eqnarray}\label{S field}&
\sum_{n_{1},..n_{d}} h(\mu_{{2\pi\over L}(n_{1},..n_{d})})  = L^{d} L^{-d}\sum_{n_{1},..n_{d}} h(\mu_{{2\pi\over L}(n_{1},..n_{d})})\nn & \sim { L^{d}\over (2\pi)^{d}} \int \D^{d}{\bf k}h(\mu_{\bf k})
\end{eqnarray}
 \begin{comment}
\bea g_{k}=\la \phi^2\ra_{\bf k}=\int \D {\bf x} e^{{i\bf x\cdot\bf k}}\la\phi({\bf x},t')\phi({\bf 0},t)\ra|_{t'\rightarrow t_{+}}\eea and similarly, using the equation of motion $\pi=\dot{\phi}$, 
  \bea h_{k}=\partial_{t}\partial_{t'}\int \D {\bf x} e^{{i\bf x\cdot\bf k}}\la\phi({\bf x},t')\phi({\bf 0},t)\ra|_{t'\rightarrow t_{+}}.\eea  \end{comment}
In other words, the field entropy {\it per unit volume} can be written as:
\begin{eqnarray}\label{S field}
{{\cal S}_{field}}\propto \int \D^{d}{\bf k}h(\mu_{\bf k})
\end{eqnarray}

Let us check the resulting symplectic eigenvalues $\mu_{k}$ for a simple case: that of a free Gaussian field. In this case we have:
\begin{eqnarray}\label{corrphi free}
g_{k,free}=
\int_0^{\infty} \D\omega 
 {1\over  {\omega^2  +k^2}}={\pi\over 2 |k|}
\end{eqnarray}
 and, 
  \begin{eqnarray}\label{corrpi free}
h_{k,free}=
\int_0^{\infty} \D\omega 
 {1\over  {\omega^2 +k^2}}k^2={\pi |k|\over 2}
\end{eqnarray}
we immediately get that the symplectic eigenvalues are all $1$: 
\bea
\mu_{k,free}=2\pi^{-1}(g_{{\bf k,free}}h_{\bf k,free})^{{1/2}}=1.
\eea
Thus, we immediately get ${{\cal S}_{field}}=0$ since in the entropy formula \eqref{Entropy formula}, $h(\mu)|_{\mu=1}=0$:  the action describing a free field does not carry any entropy. 

The last calculation may seem as a rather convoluted way of reaching the simple conclusion that the free field is in a pure state. However, the calculation serves as an important check for us that the present treatment is consistent. It is also clear how it can be adapted to more complicated situations, as studied in the next sections.  

Next, we consider the state of the field choosing a typical dielectric function $\epsilon$ to use in \eqref{corrphi} and \eqref{corrpi}. Concretely, 
we use a typical susceptibiliy of the form $\chi(\omega)={{\omega_{p}^{2}}\over (\omega_0^{2}-\omega^2-i\gamma_p \omega)}$. %(for a conductor we will add a Drude function $\chi_{c}= \chi_b+i{\omega_{c}^{2}\over \omega(\gamma_c-i \omega)}$). % (where the conductivity is given as usual by $\sigma={ne^2\over m}{1\over  (\gamma_c-i \omega)}$).% ( {\bf remark:} these definitions are in SI from jackson, and so lack the $4\pi$ factor from cgs used in my definitions here). 

Using the residue theorem, the integrals \eqref{corrphi},\eqref{corrpi} can be  expressed in terms of  the roots of the fourth order polynomial appearing in the denominator of the integrands. However the expression is rather cumbersome, but can be easily used to numerically study the symplectic eigenvalues of $\gamma$ as function $k$.
In Fig.\ref{Effective E change Wp} 
we exhibit the $k$ dependence of the effective energies $E_{k}=2\Lambda_{k}$ appearing in \eqref{rho effective homeogenous} for various values of the coupling strength of the field to the medium as represented by $\omega_{p}$.

Fig.\ref{Effective E change Wp} clearly shows, that $k$ dependence is very different from the typical linear dispersion of photon energies, showing that the state is not a thermal state of photons. It is also clear that the low momenta and high momenta asymptotics of the $E_{k}$ are quite different. In the following sections we proceed to compute the symplectic eigenvalues of the field covariance by asymptotic analysis of the integrals \eqref{corrphi},\eqref{corrpi}  in the low and high $k$ limits.

\begin{figure}
\includegraphics[scale=0.6]{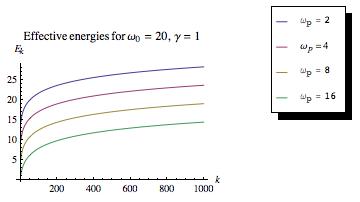}\caption{Effective energies for various values of the coupling $\omega_{p}$.}\label{Effective E change Wp}
\end{figure}

\subsection{High $k$ asymptotic behavior of symplectic eigenvalues}
To compute the cutoff dependence of the entropy for $\chi_{b}$, we consider the large $k$ asymptotics of $g_{k}, h_k$. 
Using \eqref{corrphi} at large $k$ we rewrite $g_{k}$ as:
\bea&
g_k=\int_{0}^{\infty} \frac{1}{w^2+k^2+ \omega_{p}^{2}(1-B)} \, dw\,.\eea
%where $B=\omega_{p}^{2}$, 
%$m=\omega_{0}^{2}$ and $g=\gamma_{p}$. 
where $B=\frac{\gamma_{p} w +\omega_{0}^{2}}{w^2+\gamma_{p} w +\omega_{0}^{2}}$. Noting that $  \omega_{p}^{2} B< \omega_{p}^{2}$, we can expand in series in $B$, obtaining, to lowest order in $B/(\omega^2+k^2+\omega_{p}^{2})$:
\bea&
g_k=\int_{0}^{\infty}  dw (\frac{1}{w^2+k^2+ \omega_{p}^{2}}+{\omega_{p}^{2}}\frac{1}{(w^2+k^2+{\omega_{p}^{2}})^{2}}\frac{\gamma_{p} w +\omega_{0}^{2}}{w^2+\gamma_{p} w +\omega_{0}^{2}})  +\nn &\text{higher} \,\text{orders}\sim\frac{\pi }{2} \frac{1}{\sqrt{{\omega_{p}^{2}}+k^2}}+{\omega_{p}^{2}} II\eea
with $II=\int_{0}^{\infty}  dw {\omega_{p}^{2}}\frac{1}{(w^2+k^2+{\omega_{p}^{2}})^{2}}\frac{\gamma_{p} w +\omega_{0}^{2}}{w^2+\gamma_{p} w +\omega_{0}^{2}}$. 
{let us compute II. 

Writing: 
$
\omega^2+\gamma_{p} \omega + \omega_0^{2}=\left(\omega-a_-\right)\left(\omega-a_+\right),$
where  
$a_{\pm }=\frac{-\gamma_{p}\pm \sqrt{\gamma_{p}^2-4\omega_0^{2}}}{2}$,
 there are two cases:

1)  
$
\gamma_{p}^2>4\omega_0^{2}.
$
{  In this case }
$
a_{\pm }$
\text{ are  real and negative
and carrying the integral we get }
\bea&
\text{II}=\frac{\gamma_{p} \log k }{k^4}-\frac{\gamma_{p}}{2 k^4}+\nn & \frac{-\log \left[-a_-\right] a_-^2+\log \left[-a_+\right] a_+^2}{k^4 \left(-a_-+a_+\right)}
\eea
\text{
2) }
$
\gamma_{p}^2<4\omega_0^{2}.$
In this case 
$
a_{\pm }$
{ are complex conjugates and 
we find }
\bea&
\text{II}=\frac{\gamma_{p} \log k}{k^4}-{\gamma_{p}+\gamma_{p}\log \omega_0^{2}\over 2k^{4}}+\nn & \frac{1}{k^4 \sqrt{4\omega_0^{2}-\gamma_{p}^2}}(\gamma_{p}^2-2 \omega_0^{2})  {arg}[\frac{\gamma_{p}}{2}-\frac{i}{2}\sqrt{4\omega_0^{2}-\gamma_{p}^2}]
\eea
{
In particular, if $\gamma_{p}=0$ we have:
$II=\frac{\pi\omega_0}{2 k^4}$.
%\end{comment}
Similarly, we write:
\bea&
h_k=%\int_{0}^{\infty} \left(k^2+\frac{w^{2 }\omega_{p}^{2}}{w^2+\gamma_{p} w +\omega_{0}^{2}}\right)\frac{1}{w^2+k^2+\frac{w^{2 }\omega_{p}^{2}}{w^2+\gamma_{p} w +\omega_{0}^{2}}}dw=%\nn &
%\left(k^2+B\right)g_k-B\int_{0}^{\infty} \frac{g w +m}{w^2+g w +m}\frac{1}{w^2+k^2+\frac{w^{2 }B}{w^2+g w +m}}dw=\nn & 
\left(k^2+{\omega_{p}^{2}}\right)g_k-{\omega_{p}^{2}} \text{III}\eea
where to lowest order:
$$\text{III}=\int_{0}^{\infty} \frac{\gamma_{p} w +\omega_{0}^{2}}{w^2+\gamma_{p} w +\omega_{0}^{2}}\frac{1}{w^2+k^2+{\omega_{p}^{2}}}dw$$
Carrying through the integrals we find again two cases:
1)  
$\gamma _p{}^2>4\omega _0{}^2.$ Then 
$ 
a_{\pm }$ are real and negative and
\bea
\text{III}=\frac{\log[k] \gamma _p}{k^2}-\frac{\log\left[-a_-\right] a_-^2}{k^2 \sqrt{\gamma _p{}^2-4\omega _0{}^2}}+\frac{\log\left[-a_+\right] a_+^2}{k^2 \sqrt{\gamma _p{}^2-4\omega _0{}^2}}
\eea
{2) }
$
\gamma _p{}^2<4\omega _0{}^2.$
 $a_{\pm }${ are complex conjugates.
we find:}
\bea& \text{III}=\frac{\gamma _p \log[k]}{k^2}-\frac{ 2 \gamma _p \log\left[\omega _0\right]}{2k^2}+\nn & \frac{\left(\gamma _p^2-2  \omega _0{}^2\right) \text{arg}\left[\gamma _p-i \sqrt{4 \omega _0{}^2-\gamma _p{}^2}\right]}{k^2\sqrt{4 \omega _0{}^2-\gamma _p{}^2}}\eea
{3) $g=0$.} In this case:
\bea
\text{III}=\frac{\pi  \omega_{0}}{2 k^2}\eea
%%%%
Computing to lowest order in $\omega_{p}$ the corrections $II,III$,
we find for large $k$, small $\omega_{p}$ and  $\gamma_{p}\neq 0$ 
\bea \label{large k gkhk}
g_kh_k=\frac{\pi ^2}{4}+\frac{\omega_{p}^{2}\gamma_{p} \pi  \log k}{2 k^3}.\eea
We remark that if $\gamma_{p}=0$ we get a slightly modified result:
$
g_kh_k=\frac{\pi ^2}{4}+\frac{4\pi \omega_{0} \pi ^2}{4 k^3}.
$
%Bellow 

%To compute the entropy, we estimate the symplectic eigenvalues $\mu_{\bf k}$ %using eq \eqref{corrphi},\eqref{corrpi} 
We conclude that the symplectic eigenvalues $\mu_{\bf k}$, 
to lowest order in $\omega_{p}$, behave as:
 \bea\mu_{\bf k}\sim 1+{\omega_{p}^{2}\gamma_{p}\log k\over \pi}+..\eea for $k\gg1$.

\subsection{Low $k$ behavior of symplectic eigenvalues}
Studying the low $k$ behavior requires slightly more care in the expansion.
Here we find that the leading behavior for the field correlators at low $k$ is given by
\bea
g_{k}\sim\frac{\pi }{2 k\sqrt{1+ \frac{\omega_{p}^{2}}{\omega_{0}^{2}}}}.\eea

To do so, we start by rescaling $w$ in the integral \eqref{corrphi}:
\bea& g_{k}=\int_0^{\infty }\frac{1}{w^2+k^2+\frac{w^{2 }{\omega_{p}^{2}}}{w^2+\gamma_{p} w +\omega_{0}^{2}}}dw= \nn &\frac{1}{k}\int_0^{\infty }\frac{1}{ 1+x^2+x^2\frac{{\omega_{p}^{2}}}{\omega_{0}^{2}}\left(1-\frac{k x(k x+\gamma_{p})}{k^2x^2+\gamma_{p} k x +\omega_{0}^{2}}\right)}d x.
\eea
We continue to split the integral into two intervals $(0,k^{-1})$ and $(k^{-1},\infty)$, and separately approximating each term:
\bea&
\frac{1}{k}\int_0^{1/k}\frac{1}{ 1+x^2+x^2\frac{{\omega_{p}^{2}}}{\omega_{0}^{2}}(1-\frac{k x(k x+g)}{k^2x^2+\gamma_{p} k x + \omega_{0}^{2}})}d x+\nn & \frac{1}{k}\int_{1/k}^{\infty }\frac{1}{ 1+x^2+x^2\frac{{\omega_{p}^{2}}}{\omega_{0}^{2}}(1-\frac{k x(k x+\gamma_{p})}{k^2x^2+\gamma_{p} k x +\omega_{0}^{2}})}d x\nn & \sim \frac{1}{k}
\int_0^{1/k}\frac{1}{ 1+x^2+x^2\frac{{\omega_{p}^{2}}}{\omega_{0}^{2}}}d x+\frac{1}{k}\int_{1/k}^{\infty }\frac{1}{ 1+x^2}d x
\eea
{for small k we get:
\bea
\frac{1}{k}\int_{1/k}^{\infty }\frac{1}{ 1+x^2}d x\to 1
\eea
\text{
and:}
\bea&
\frac{1}{k}\int_0^{1/k}\frac{1}{ 1+x^2+x^2\frac{{\omega_{p}^{2}}}{\omega_{0}^{2}}}d x=\frac{\text{arctan}\left[\frac{1}{k}\sqrt{1+\frac{\omega_{p}^{2}}{\omega_{0}^{2}}} \right]}{k\sqrt{1+\frac{{\omega_{p}^{2}}}{\omega_{0}^{2}}}}\to \nn &  \frac{\pi }{2 k\sqrt{1+ \frac{\omega_{p}^{2}}{\omega_{0}^{2}}}    }\eea
a similar treatment shows that:
\bea
h_{k}\sim \frac{\pi \omega_{p}^{2} }{\sqrt{4\omega_{p}^{2} -\gamma_{p}^2+4 \omega_{0}^{2}}}+o(k).
\eea
Finally, we find that the symplectic eigenvalues diverge at $k\rightarrow 0$ since
\bea
\sqrt{g_kh_k}\sim \frac{\pi }{\sqrt{2 k}}\left(\frac{ \omega_{0}^{2} ({\omega_{p}^{2}})^2 }{( \omega_{0}^{2}+{\omega_{p}^{2}}) \left(4\omega_{p}^{2}-\gamma_{p}^2+4 \omega_{0}^{2}\right)}\right)^{1/4}
\eea
diverges as 
$\frac{1}{\sqrt{2k}}$ for small $k$. The entropy function $h(k)$ for small $k$s diverges as $h(k)\sim -log(k)$. However, integration $d^{D}k$ is finite for any dimension and so there are no infra-red divergences in the total entropy per unit volume.

\subsection{Interpretation of the field reduced density matrix}
Having evaluated the symplectic eigenvalues of the field covariance matrix we can  write
the density matrix of the field as \eqref{Effective rho normal}
\bea 
\rho_{}=Z^{-1}e^{-\sum_{k} 2\Lambda_{k}{a^{+}}_{k}a_{k}}
~;~2\Lambda_{k}=\log{\sqrt{\frac{4}{\pi ^2}g_kh_k}+1\over \sqrt{\frac{4}{\pi ^2}g_kh_k}-1}
\eea
As mentioned in the introduction, the first natural question to ask is: is the state of the field $\rho$ thermal? This is certainly a natural possibility suggested by considering the bulk material as a thermal bath for the field. In fact, one may always write the state $\rho$, formally,  as thermal, i.e.: \bea \rho=Z^{-1} exp(-\beta H_{eff}),\eea
for some suitable operator $H_{eff}$. However, to actually interpret the state as thermal, we would like $H_{eff}$ to represent a reasonable, local, physical Hamiltonian. For the Gaussian states we can write  $H_{eff}$ explicitly as:
\bea H_{eff}=\int\D{\bf x}\D{\bf x}'  \hat{u}({\bf x}-{\bf x}' )a^{+}_{{\bf x}}a_{{\bf x}'}\eea 
where 
\bea \hat{u}=2\beta^{-1} \int\D{\bf k} \Lambda_{\bf k}e^{i {\bf k}\cdot({\bf x}-{\bf x}' )}.\eea
The locality properties of this Hamiltonian thus depend on the locality of the Fourier transform. We find that generically
the Fourier transformed $\hat{u}$ gives us a non-local $H_{eff}$. 

Alternatively, we may interpret $2\Lambda_{\bf k}$ as $2\Lambda_{\bf k}=\beta_{k}\omega _k$ where $\omega _k$ are the photon energies in a free homogenous space, without the interaction with the material, i.e. 
\bea \label{effective temp}
\beta _k\omega _k=2\Lambda_k \sim \log \left(\frac{\sqrt{\frac{4}{\pi ^2}g_kh_k}+1}{\sqrt{\frac{4}{\pi ^2}g_kh_k}-1}\right)
\eea
As is clearly shown in fig.\ref{Effective E change Wp}, if we take $\omega_k\propto |k|$ we find that different momenta ${\bf k}$ feel different effective temperatures. 

In the infra-red limit expanding this expression for small $k$ using we find the leading, small $k$ behavior:
\bea&
\beta _k \omega _k\sim \nu \sqrt{k}~~;  \nn & \nu \equiv\sqrt{2}\left(\frac{\omega_0^2 ( \omega_p^2)^2 }{(\omega_0^2+ \omega_p^2) \left(4\omega_p^2-\gamma_p^2+4 \omega_0^2\right)}\right)^{-1/4}
\eea
If we assume $\omega _k\propto |k|$ we conclude that the effective temperature of this $k$ mode is:
\bea
T_k\equiv{k\over \beta_{k}}\propto {\sqrt{k}}~~~as~~~~k\rightarrow 0.
\eea
What is the energy and number of occupied soft modes per unit volume up to a given $k_m$? 
We show that these are proportional to $k_m^{d+1/2}$ and $k_m^{d-1/2}$, respectively, are finite and small. Indeed, since the occupation of a mode k is given by:   
\bea n_k=\frac{1}{e^{\beta _k \omega_k}-1}\sim  \frac{1}{\nu  \sqrt{k}}\eea
The expected occupation number for modes with $k<k_{\min }$ is thus given by:
\bea
\text{   }N_{\text{kmin}}=\int_{k<k_{\min }}d^dk n_k\propto  \int_0^{k_{\min }}\text{dk} \frac{k^{d-1}}{\nu \sqrt{k}}=\frac{k^{d-1/2}}{\nu  \left(d-\frac{1}{2}\right)}
\eea
The energy, assuming 
$\omega_k\propto k$, 
{   is: }
\bea&
E\left(k<k_{\min }\right)\sim \int_{k<k_{\min }}d^dk n_k k\propto   \nn & \int_0^{k_{\min }}\text{dk} \frac{k^d}{\nu \sqrt{k}}=\frac{k^{d+1/2}}{\nu  \left(d+\frac{1}{2}\right)}
\eea
Finally, the number variance of modes up to  $k_m$ is computed as:
\bea&
\delta N^2\left(k<k_{\min }\right)=\int _{k<k_{\min }}d^dk n_k \left(1+n_k\right)\propto  \nn & \int _0{}^{k_{\min }}\text{dk} k^{d-1} \frac{1}{\mu ^2k}=\nn & \int _0{}^{k_{\min }}\text{dk} k^{d-2} \frac{1}{\mu ^2}= \begin{cases}
 \frac{1}{\mu ^2}\log  \frac{k_{\min }}{\varepsilon_{IR}} & d=1 \\
 \frac{1}{\mu ^2}\frac{\left(k_{\min }\right){}^{d-1}}{d-1} & d>1
\end{cases}.\eea  
Note that for $d=1$ we find an  infra-red divergence: $\la \delta N^2\ra=-\log \varepsilon_{IR}$, where $\varepsilon_{IR}$ is an infra-red cutoff, inversely proportional to the system size.

\subsection{Cutoff dependence of the entropy}
In this section, we compute the total field entropy, and show that it suffers from a UV divergence. This result is not surprising, since, in principle, entanglement can get contributions from all momentum scales.

Using \eqref{S field}, the full quantum von-Neumann entropy of the field per unit volume is given by the momentum integrals of $h(\sqrt{\frac{4}{\pi ^2}g_kh_k})$. The large momentum behavior of the integrand is obtained from  \eqref{large k gkhk} to be:
\bea& \label{h large k}
h(\sqrt{\frac{4}{\pi ^2}g_kh_k})\sim \frac{\omega_{p}^{2} \gamma_{p}  \log k(1+3  \log k-\log \frac{\omega_{p}^{2} \gamma_{p}  \log k}{2 \pi })}{2 \pi  k^3}
\eea
for a transparent medium ($\gamma_{p}<<1$) we consider additional terms and find:
\bea\label{h large k no gamma}
h(\sqrt{\frac{4}{\pi ^2}g_kh_k})\sim \frac{{\omega_{p}^{2}}  {\omega_{0} }(1+3 \log k-\log  \frac{\omega_{p}^{2}  {\omega_{0} }}{4})}{4 k^3}
\eea
{using the integrals  }
$
\int \frac{\log[k]}{k}=\frac{\log[k]^2}{2}$
{ and }
$\int \frac{\log^2[k]}{k}=\frac{\log[k]^3}{3}$,
we can estimate the integration of the expressions \eqref{h large k}, \eqref{h large k no gamma} over $k$ in 3d, to find the entropy per unit volume
\bea\label{Results entropy cutoff}&
%{{\cal S}_{field}\over Vol}=-{1\over Vol}\Tr \rho_{field}\log\rho_{field}\sim \nn & 
{\cal S}_{field}\propto\Big\{
\begin{array}{cc}
\omega_{p}^{2}\gamma_p \log(\Lambda)^{3}  &     \\
\omega_{p}^{2}\omega_{0}\log(\Lambda)^{2}  &  \gamma_{p}=0  \\
 0 &    plasma\,\, model
\end{array}, 
\eea
where $\Lambda$ is a high momentum (UV) cutoff. %Note that in \eqref{Results entropy cutoff} for simplicity we summed over the bound and free charge contribution in $\epsilon$, which, strictly speaking is only valid at the dilute limit.

Note that in obtaining  \eqref{Results entropy cutoff} is obtained to lowest order in $\omega_{p}$. However, this approximation is justified for our purposes since it  becomes exact at the large $k$ limit which we are studying. Indeed, numerically, the approximations used in \eqref{Results entropy cutoff} actually recover the correct cutoff dependence even for large values of $\omega_{p}$, since the dielectric response decays at large $k$ values. 
%This highlights the different roles of generating quantum entropy, where the dissipative terms, describing interaction with additional degrees of freedom, are dominant. However, even when they are absent we still have entanglement between the field and the dielectric.

It is interesting to observe the special place of the  {\it "pure plasma"} limit response function. %({\bf Real or imaginary frequencies?}). 
We can easily understand the result \eqref{Results entropy cutoff} as follows: Substituting the plasma permittivity limit form $\chi=- { \omega_{c}^2\over \omega^2}$ in the action \eqref{action}, in a homogenous space, we see that the role of $\chi$ is similar to producing a mass term for $\phi$. Thus, the resulting action is consistent with a Hermitian field Hamiltonian, and as such, at zero temperature, to a pure state. One can also understand the vanishing of entropy for pure plasma as follows: Consider a slab of material, with a pure plasma form for the dielectric response. Such a material will have no losses: A pure plasma system will be completely transparent to radiation above the plasma frequency, and completely reflecting at lower frequencies. The field will have a finite mass inside the region occupied by the plasma and will not generate entropy.
%({\bf say something about degeneracies?}). 
It is interesting that the use of a pure plasma in the computations of Casimir energy and Entropy has been at the heart of a recent debate \cite{bezerra2004violation,decca2005precise,brevik2005temperature,Sushkov:2011fk}. We do not make any claims regarding the debate, but note that the distinction between the Casimir entropy in the two models is manifested in the full quantum entropy computed herein. 

We remark that it is possible to incorporate the high momentum cutoff more naturally by taking into account the spatial dispersion in  $\epsilon$. While different functions may have different asymptotic properties, we expect to get similar behavior. Indeed, if we take, for concreteness sake, the simplest extension of the previous treatment, we can use  (see e.g. \cite{klingshirn2005semiconductor}).
\begin{eqnarray}
\epsilon (i \omega
   ,k)=\epsilon
   _0\left(1+\frac{f}{
   \text{A k}^2 +\gamma  \omega +\omega
   ^2+\omega _0^2}\right)
\end{eqnarray}
where the expression is valid for small $k\ll \pi/a$ where $a$ is the interatomic distance. One can easily check that this form doesn't change much the cutoff dependence.

\section{Distance dependent entropies}

Since the entropy ${{\cal S}_{field}}$ is UV divergent, it is natural to ask, in analogy with the Casimir effect, what is the distance dependence of the entropy of interaction with two distinct bodies $A$ and $B$, and is the distance dependent part of it is UV finite. 
To answer such questions, the  ``Casimir EE'' was defined in \cite{klich2012prlentanglement}  as
%S_{{\phi}}(A,B)=dist.~depend.~part~of~-\Tr \rho_{\phi}\log\rho_{\phi}=
\bea{\cal S}_{{R,\alpha}}(A,B)={\cal S}_\alpha(A\cup B)-{\cal S}_\alpha(A)-{\cal S}_\alpha(B),\eea 
where ${\cal S}_{{\alpha}}(M)$ is the $\alpha-$Renyi entropy of a field interacting with a body $M$
described by $\epsilon=\epsilon_{0}+ \chi_{M}$, where the susceptibility $ \chi_{M}=0$ everywhere outside $M$. The situation is illustrated in Fig.\ref{objects}

\begin{figure}
\includegraphics[scale=0.4]{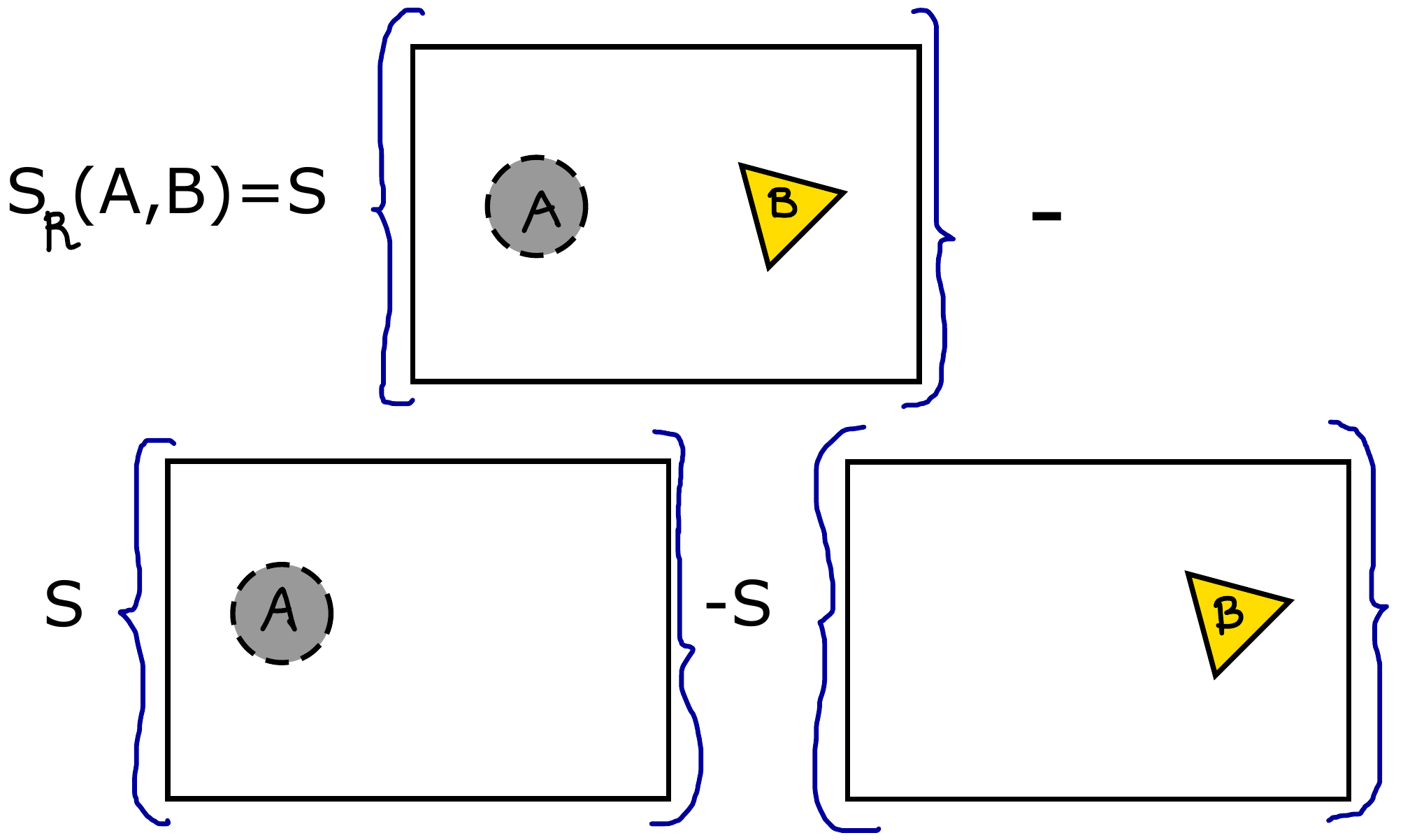}\caption{A Casimir entanglement entropy.}\label{objects}
\end{figure}

%The entropy of the field may be quantified in various ways. Here we use as a measure the von-Neumann of the field. Thus for a pair of bodies $A$ and $B$ we define
%\begin{eqnarray}
%S_{{\phi}}(A,B)=dist.~depend.~part~of~-\Tr \rho_{\phi}\log\rho_{\phi}
%\end{eqnarray}
%where $\rho_{\phi}$ is the density matrix of the field $\phi$, in the presence of a body $A$ and $B$. 

%For the full electromagnetic problem this calculation is challenging. As we will see, even for a Gaussian, scalar field it is very hard to make progress analytically and we have to resort to numerics. 

The following few remarks regarding ${\cal S}_{{R}}$ are important to note:

1. ${\cal S}_{{R}}$ is distinct from the more familiar 
Casimir entropy. Casimir entropy is defined as ${\cal S}_{C}(A,B)=-\lim_{T\rightarrow 0}\partial_{T}F_{C}$, where $F_{C}$ is the Casimir free energy, obtained by subtracting all distance independent terms from the free energy of the EM field in the presence of bodies $A,B$ or boundary conditions. 

2. ${\cal S}_{{R}}$ is different from the "relative entropy" of probability theory, as we are comparing different systems, and not merely different statistical information about the same system.

3. It is important to note that, while the sub-additivity of von-Neumann entropy \cite{araki1970entropy} shows that the entropy ${\cal S}$ of a field is always larger than the total entropy of the combined system of field matter, this is no longer evident once subtractions are taking place in order to define ${\cal S}_{R}$.% and ${\cal S}_{C}$.

The relevance of Casimir entropy ${\cal S}_{C}$ to understanding thermal corrections of the Lifshitz formula has been pointed out in many papers (see e.g. \cite{bezerra2004violation,Geyer:2005ap,bordag2010casimir}), where it was noticed that as $T\rightarrow 0$, ${\cal S}_{C}$ may not go to zero when using the Drude model, as would be expected by the Nernts theorem. It is quite interesting to note that while the Casimir entanglement entropy ${\cal S}_{{R}}$ is distinct from ${\cal S}_{C}$, a similar behavior is observed in \eqref{Results entropy cutoff}.
${\cal S}_{C}$ has a clear thermodynamic meaning, especially at high temperatures, where the Casimir force is entirely entropic \cite{feinberg2001casimir}.
Indeed, at high temperatures we expect ${\cal S}_{R}={\cal S}_{C}$, as most of the field entropy will be thermal (Technically, the relevant Green's function gets its major contribution from the $\omega=0$ Matsubara pole).

\section{Formulas for the distance dependent Von Neumann and Renyi entropies}
In this section we describe the derivation of an abstract formula presented in \cite{klich2012prlentanglement} for the Casimir entanglement entropy. In addition, we generalize the formula to also describe arbitrary Renyi entropies. It was shown in that \cite{klich2012prlentanglement} :
\bea \label{S relative}&
{\cal S}_{R}(A,B)=-\frac{1}{\pi }\int _{1/2-\text{i$\infty $}}^{1/2+\text{i$\infty $}}dx\frac{\log\left[\frac{\sqrt{x}+\frac{1}{2}}{\sqrt{x}-\frac{1}{2}}\right]}{2 \sqrt{x}}\Tr\log\nn &\left(1-\frac{1}{1-K_A}\left(K_AK_B+K_{\text{AUB}}-K_A-K_B\right)\frac{1}{1-K_B}\right)\eea
where $
K_A= \frac{1}{x-1/4+i s}\left(\Gamma_{A}-1/4\right)
$, and similar expressions hold for $K_{B},K_{AUB}$. Here $\Gamma={\cal G}_{A}{\cal H}_{A}$, where ${\cal G}_{A}$ and ${\cal H}_{A}$ are field and field momentum two point functions, respectively, in the presence of body $A$. 

In this section we repeat the derivation of \eqref{S relative} and generalize the expression to general Renyi entropies:
\begin{eqnarray}\label{SRalpha general}&
S_{R,\alpha}=-\frac{1}{\pi }\int _{1/2-\text{i$\infty $}}^{1/2+\text{i$\infty $}}\D x \frac{\left(\left(2 \sqrt{x}+1\right)^{\alpha -1}-\left(2 \sqrt{x}-1\right)^{\alpha -1}\right) \alpha }{\left(\left(2 \sqrt{x}+1\right)^{\alpha }-\left(2 \sqrt{x}-1\right)^{\alpha }\right) \sqrt{x} (\alpha -1)}\Tr\log\nn &\left(1-\frac{1}{1-K_A}\left(K_AK_B+K_{\text{AUB}}-K_A-K_B\right)\frac{1}{1-K_B}\right)
\end{eqnarray}
%\new 
The expression \eqref{S relative}, is similar to the TGTG formulas in it's form: an integral over the TrLog of a combination of Green's functions. However, it differs from such formulas in three major aspects: 

1) The integration variable $x$ is not a frequency variable, but rather an auxiliary spectral variable, 

2) The presence of the term $K_{AUB}$ does not allow for full separation into local object properties and free propagators. 

3) The non-analyticity of the integrand at $x=1/4$. All of these make the formula harder to use than the TGTG formulas. Nevertheless, it can be used as a starting point for various expansions when the bodies are  weakly entangled with the field, so that $||\Gamma_{A}-1/4||\ll 1$.

The derivation bellow of eq. \eqref{S relative} follows and adapts the approach of \cite{ kenneth2008casimir} to the scattering formalism for Casimir energies.

To compute the entropy, we first need to find an expression for the density of  symplectic eigenvalues of the covariance matrix $\gamma$. In the absence of $\la\phi\pi\ra$ correlations, these symplectic eigenvalues (up to a factor 2 in the definition of the covariance matrix eq. \eqref{Covariance Mat}) are related to the square roots of eigenvalues of $\Gamma$. 

Thus, we first find a convenient representation to the density of density of states of $\Gamma={\cal G}{\cal H}$. Note that $\Gamma$ is not Hermitian, however, since ${\cal G}$ and ${\cal H}$ are positive Hermitian matrices, it has the same spectrum as ${\cal G}^{1/2}{\cal H}{\cal G}^{1/2}$, and one may safely use the formulas below.
Consider the representation:
\bea & \delta (E-x)=\frac{1}{\pi }\text{Im} \frac{1}{E-x+i 0}=\nn &
\frac{1}{\pi }\text{Im} \partial _E \text{Log}(E-x+i 0)\eea
Then we have: relative density of states of a Hermitian operator X as:
\bea &\label{relative rho}
\delta \rho_{\Gamma} (E)\equiv \rho_{\Gamma}(E)-\rho_{\Gamma_{0}}(E)
=\nn &
\frac{1}{\pi }\text{Im} \partial _E\left[ \text{TrLog}(E-\Gamma +i 0)-\text{TrLog}\left(E-\Gamma_0+i 0\right) \right]=\nn &\frac{1}{\pi }\text{Im} \partial _E\text{TrLog}\left(({E-\Gamma +i 0})({E-\Gamma_0+i 0})^{-1}\right)\nn & =\frac{1}{\pi }\text{Im} \partial _E\text{TrLog}\left(1+\left(\Gamma_0-\Gamma\right)\frac{1}{E-\Gamma_0+i 0}\right)\eea
Denoting:
$G_0=
\frac{1}{E-\Gamma_0+i s}$ and 
 $ D=
\left(\Gamma-\Gamma_0\right)$,
we can write \eqref{relative rho} as
$$
\delta \rho_{\Gamma} (E)=\frac{1}{\pi }\text{Im} \partial _E\text{TrLog}\left(1-G_0 D \right)$$
To compute
\bea{\cal S}_{{R}}(A,B)={\cal S}(A\cup B)-{\cal S}(A)-{\cal S}(B),\eea
we need the relative densities:
\bea\label{Relative density of states} &
\delta\rho_{R}\equiv\delta \rho_{\Gamma_{AUB}}-\delta \rho_{\Gamma_{A}}-\delta \rho_{\Gamma_{B}}= \nn &
\frac{1}{\pi }\text{Im} \partial _E[\text{Tr log}\left(1-G_0 D_{AUB} \right)-\text{Tr log}\left(1-G_0 D_{A} \right)\nn & -\text{Tr log}\left(1-G_0 D_{B} \right)].
\eea
Here $D_{A}$ is computed using $\Gamma_{A}={\cal G}_A{\cal H}_A$, where the correlations ${\cal G}_A,{\cal H}_A$ are computed for the field in the presence of body $A$, and similarly for $D_{AUB},D_{B}$. We combine the terms using the following identity:
%The $TGTG$ stuff doesnt work in our case since $D_{\text{AUB}}\neq D_A+D_B. $ but we can write:
\begin{eqnarray*}
&
\frac{1}{1-G_0D_A}\left(1-G_0D_{\text{AUB}}\right)\frac{1}{1-G_0D_B}=\nn & \frac{1}{1-G_0D_A}((1-G_0D_A)\left(1-G_0D_B\right)- \nn & G_0D_AG_0D_B-G_0(D_{\text{AUB}}-D_A-D_B))\frac{1}{1-G_0D_B}= \nn & 1-\frac{1}{1-G_0D_A}\left(G_0D_AG_0D_B+D_{\text{AUB}}-D_A-D_B\right)\frac{1}{1-G_0D_B}
\end{eqnarray*}
Defining:
$$K_A=G_0D_A=\frac{1}{E-\Gamma_0+i s}\left(\Gamma_{A}-\Gamma_0\right)$$
we obtain
\begin{eqnarray*}&\label{relativeDOS}
\delta \rho (E)_{R}=\frac{1}{\pi }\text{Im} \partial _E\text{TrLog}\Big(1-\nn &\frac{1}{1-K_A}\left(K_AK_B+K_{\text{AUB}}-K_A-K_B\right)\frac{1}{1-K_B}\Big)\end{eqnarray*}

\begin{comment}{In our case we are taking }
$$
K_A= \frac{1}{E-G_0 {\cal H}_0+i s}\left(G_AH_A-G_0 {\cal H}_0\right)
$$
Note that, for example 
$K_A+K_B-K_{\text{AUB}}$
{ is in fact second order as well, as it involves terms of the form }
$T_A+T_B-T_{A U B}$. Also, 
things get simpler, since we have $G_0 {\cal H}_0=1/4$. Thus we have:
$$
K_A= \frac{1}{E-1/4+i s}\left(G_AH_A-G_0 {\cal H}_0\right)
$$
\end{comment}
We now use this spectral density to compute the entropy using the eigenvalues of the covariance matrix, eq. \eqref{Entropy formula},
\begin{eqnarray*}&
{\cal S}=\sum_{i} h(\mu_{i}) ~;~ h(\mu)={\mu +1\over 2}\log{\mu +1\over 2}-{\mu-1\over 2}\log{\mu-1\over 2}
\end{eqnarray*} 
The lowest eigenvalue ${\cal G}{\cal H}$ is always larger or equal to 1/2, by the uncertainty relations. Thus we may write
\bea
S=\int _{1/2}^{\infty }h\left(2\sqrt{x}\right) \delta \rho (x)_R dx \eea
Integrating by parts and moving the contour integration to the imaginary axis, 
we write this expression as:
\begin{eqnarray*}&
S=-\frac{1}{\pi }\int _{1/2-\text{i$\infty $}}^{1/2+\text{i$\infty $}}\D x \frac{\text{Log}\left[\frac{\sqrt{x}+\frac{1}{2}}{\sqrt{x}-\frac{1}{2}}\right]}{2 \sqrt{x}}\text{TrLog}\nn &\left(1-\frac{1}{1-K_A}\left(K_AK_B+K_{\text{AUB}}-K_A-K_B\right)\frac{1}{1-K_B}\right)
\end{eqnarray*}
which is eq. (11).

We can also extend this formula to cover Renyi entropies \eqref{Renyi entropy def}. Using the formula for the density of states \eqref{Relative density of states} together with the expression for the Renyi entropy in terms of the symplectic eigenvalues \eqref{Renyi entropy using symplectic eigenvalues}, we find:
\bea&
S_{R,\alpha}=\nn & -\frac{1}{\pi }\int _{1/2-\text{i$\infty $}}^{1/2+\text{i$\infty $}}(\partial _Eh_{\alpha }(2\sqrt{E}))\text{Tr} \text{Log}(1- \nn &{1\over1-K_A}(K_AK_B+K_{\text{AUB}}-K_A-K_B)\frac{1}{1-K_B})dE
\eea
yielding \eqref{SRalpha general}.
In particular, the second Renyi entropy is given by:
\begin{eqnarray}\label{Renyi 2 relative} &
S_{R,2}=-\frac{1}{\pi }\int _{1/2-\text{i$\infty $}}^{1/2+\text{i$\infty $}}\D x {1\over 2x}\Tr\log(1- \nn &\frac{1}{1-K_A}\left(K_AK_B+K_{\text{AUB}}-K_A-K_B\right)\frac{1}{1-K_B}).
\end{eqnarray}

To relate this form to TGTG formulas, we recall that in such formulas, the relative density of states of the electromagnetic field interacting  with for two bodies $A,B$,  through dielectric susceptibilities $\chi_{A}(\omega), \chi_{B}(\omega)$ at frequency $\omega$ is expressed (before Wick rotation) as:
\bea
{1\over \pi}Im\partial_{\omega}\text{Tr log}(1-T_{A}G_{0}T_{B}G_{0})
\eea
where $T_{A}= \omega^{2}\chi_{A}{1\over 1-G_{0} \omega^{2}\chi_{A}}$ are the Lippman-Schwinger operators of the problem, and $G_{0}$ are free propagators. 

Choosing $K_A\frac{1}{1-K_A}$ to play the role of $T_{A}G_{0}$ and $K_B\frac{1}{1-K_B}$ the role of $T_{B}G_{0}$, The density of states \eqref{relativeDOS} may be written as:
\begin{eqnarray*}&
\delta \rho (E)_{R}=\nn &\frac{1}{\pi }\text{Im} \partial _E\text{TrLog}(1-T_{A}G_{0}T_{B}G_{0}\nn & -\frac{1}{1-K_A}\left(K_{\text{AUB}}-K_A-K_B\right)\frac{1}{1-K_B}).\end{eqnarray*}
We observe, however, the appearance of an additional $\left(K_{\text{AUB}}-K_A-K_B\right)$ term, which is not-separable into a product of correlators of the separate bodies.

\section{Representation of the correlation functions in terms of Lippmann-Schwinger operators}
In the computation of the correlation functions bellow, we will use extensively the representation of correlation functions in terms of {\it Lippmann-Schwinger} operators.
{For a body A, we define the Lippmann-Schwinger operator }
$T_A$ at imaginary frequency $i \omega$  by 
\bea\label{TA definition} T_{A}= \omega^{2}\chi_{A}{1\over 1+  g_{0}\omega^{2}\chi_{A}(i\omega)}\eea
where $g_{0}$ are free propagators. In particular, for a scalar field we take $g_0={1\over -\Delta+\omega^2}$.

The Lippmann-Schwinger operator $T_A$ is { related to }the green's function $g_A$ , 
\bea
\la x|  g_A(\omega )|x'\ra=\la x|\frac{1}{-\Delta +\omega ^2\epsilon _A(i \omega )} |x'\ra
\eea
{by the operator equation:  }
\bea
g_A(\omega )=g_0(\omega )+g_0(\omega )T_Ag_0(\omega )
\eea
The eq. \eqref{corrphi} for the $\la \phi \phi \ra$ correlation functions written in a general basis (i.e. without assuming translational invariance) is then:
\bea& \label{G in terms of w integral of T} {\cal G}_A=\int_0^\infty \text{d}\omega g_A=\int_0^\infty  \text{d}\omega\left(g_0+g_0T_Ag_0\right)= \nn &{\cal G}_0+\int_0^\infty  \text{d}\omega g_0T_Ag_0\eea
Similarly, as in \eqref{corrpi}, the field momenta correlation functions are encoded by 
\bea\label{H in terms of w integral of T}
{\cal H}_A={\cal H}_0-\int \text{d}\omega \omega ^2g_0T_Ag_0.
\eea
(in the expressions above, and what follows we omit the $\omega$ dependence in $g_{0}$).  
We note that in the continuum, ${\cal G}_0,{\cal H}_0$ are diagonal in momentum, with matrix elements given by ${\pi\over 2 |k|}\delta(k-k')$ and ${\pi |k|\over 2}\delta(k-k')$ respectively (as obtained in eq. \eqref{corrphi free},\eqref{corrpi free}).

Let us recall some general properties of $T_A$  on the imaginary frequency axis \cite{kenneth2008casimir}.  Below we will repeatedly use that, as a consequence of the Kramers-Kronig relations, combined with the assumption of equilibrium we have $\chi_{A}(i\omega)>0$. In it is known that $\chi_{A}(i\omega)$ is real and decaying as $\omega^{-2}$ as $\omega\rightarrow\infty$.

We have the following properties:

1. $T_{A}\psi=0$ for any $\psi$ for which vanishes on $A$. This is established by rewriting  $T_{A}$  as 
\bea\label{TA definition} T_{A}=\sqrt{ \omega^{2}\chi_{A}}{1\over 1+\sqrt{ \omega^{2}\chi_{A}} g_{0}\sqrt{ \omega^{2}\chi_{A}}}\sqrt{ \omega^{2}\chi_{A}}\eea

2.
$T_A$ { is a positive operator, i.e. $\langle \psi |T_A|\psi \rangle>0$ for any $\psi$ in the Hilbert space $T_A$ acts on (square integrable function supported on the region A).

And,

3. 
\bea \label{T smaller than chi} T_A<\omega ^2\left(\epsilon _A(i \omega )-1\right)\eea as operators, 
i.e. for any vector $\psi$, 
$\langle \psi |T_A|\psi \rangle <\langle \psi |\omega ^2\left(\epsilon _A(i \omega )-1\right)|\psi \rangle$.

In the next section we use these properties in our analysis of the distance dependence of the entropy ${\cal S}_{R}$.

\section{Large distance expansion of the Renyi entropy $S_{2,R}$}
In this section, we study the behavior of the relative entanglement at large distances. Consider two bodies with a dielectric function $\epsilon_{A}(i\omega)=1+{\omega_{pA}^{2}\over \omega^{2}+ \omega_{0}^{2}}$ (and similar expression for body $B$), and volumes $V_{A},V_{B}$.

Our main result is that at large separation the Renyi entropy of the field is the sum of the separate body entropies, with the correction decaying as:
\bea\label{R4 asymptotic}
S_{2,R}= -\omega _{{pA}}^2\omega _{{pB}}^2V_AV_B\frac{2\pi ^4}{\omega_{0}^2R^4}+O({1\over R^{6}})
\eea
Note that the power law differs from the typical power law of $R^{-7}$ appearing in the Casimir-Polder interaction. 

To find this result we first write the 2-Renyi entropy \eqref{Renyi2 explic} as:  \bea
S_2=\Sigma  \log  \mu  =\frac{1}{2}\text{Tr} \log  (1+\delta \Gamma )
\eea
{Thus the relative $S_{2,R}$ is given by:}
\bea&\label{S2 relative}
S_{2,R}=\frac{1}{2}\text{Tr} \log  \left(1+\delta \Gamma _B\right){}^{-1}\left(1+\delta \Gamma _{\text{AUB}}\right) \left(1+\delta \Gamma _A\right){}^{-1}\nn &=\frac{1}{2}\text{Tr} \log  \Big[1+ \nn &  \frac{1}{1+\delta \Gamma _B}\left(\delta \Gamma _{\text{AUB}}-\delta \Gamma _A-\delta \Gamma _B-\delta \Gamma _A\delta \Gamma _B\right) \frac{1}{1+\delta \Gamma _A}\Big]
\eea
and in particular, in the long distance expansion, we expect:
\bea
||\delta \Gamma _{\text{AUB}}-\delta \Gamma _A-\delta \Gamma _B-\delta \Gamma _A\delta \Gamma _B||\to 0
\eea
and we can approximate:
\bea\label{Renyi2 second order}
S_{2,R}\sim \frac{1}{2}\text{Tr} \left(\delta \Gamma _{\text{AUB}}-\delta \Gamma _A-\delta \Gamma _B-\delta \Gamma _A\delta \Gamma _B\right)
\eea
In calculations of \eqref{Renyi2 second order}, we have several different kinds of terms.
We concentrate on the so called ``dilute limit ''  where it is assumed $\omega^{2}\chi(i\omega)\ll 1$. In this case we can use the approximation $T_{A}\sim \omega^{2}\chi_{A}(i\omega)$.

To lowest order in $\chi_{A},\chi_{B}$, we have the following terms:
\bea & \label{S2R lowest} S_{2,R}\sim  \frac{1}{2}\text{Tr}  ( {\cal H}_0 ( \delta{\cal G}_{\text{AUB}}-\delta{\cal G}_A- \delta{\cal G}_B )+\nn & ( \delta{\cal H}_{\text{AUB}}- \delta{\cal H}_A- \delta{\cal H}_B ) {\cal G}_0- \nn &  {\cal H}_0 \delta{\cal G}_A {\cal H}_0 \delta{\cal G}_B-  \delta{\cal H}_A{\cal G}_0 \delta{\cal H}_B{\cal G}_0 ) )\eea

It turns out that the leading contribution is obtained from:
\bea\label{R4 asymptotic contribution}
\Tr {\cal H}_0 \delta{\cal G}_A {\cal H}_0 \delta{\cal G}_B+\delta{\cal H}_A{\cal G}_0 \delta{\cal H}_B{\cal G}_0=O\left(R^{-4}\right)
\eea
The calculation goes as follows. {Explicitly, using \eqref{corrphi free},\eqref{corrpi free} ,\eqref{G in terms of w integral of T} and \eqref{H in terms of w integral of T} :}
\bea & 
\Tr {\cal H}_0 \delta{\cal G}_A {\cal H}_0 \delta{\cal G}_B+\delta{\cal H}_A{\cal G}_0 \delta{\cal H}_B{\cal G}_0=O\left(R^{-4}\right)=\nn &  \int \text{d}\omega\text{d$\omega $}'d^3k \la k|{\cal H}_0g_0T_Ag_0{\cal H}_0g_0T_Bg_0+ \nn & \omega ^2\omega '^2{\cal G}_0g_0T_Ag_0{\cal G}_0g_0T_Bg_0|k\ra=\nn & \int \text{d}\omega\text{d$\omega $}'\text{  }d^3k d^3q\text{   }\frac{1}{k^2+\omega ^2}T_A(k,q,\omega )\frac{1}{q^2+\omega ^2}\times \nn & \frac{1}{q^2+\omega '^2}T_B(q,k,\omega ')\frac{1}{k^2+\omega '^2}\left(\frac{\omega ^2\omega '^2}{|k||q|}+|k||q|\right)
\eea
Now consider the effect of shifting the object $B$ by a vector ${R}\hat{n}$ to $B_{R}=\{x:x-{R}\hat{n}\in B \}$ . 
%$
%x\to x+R
%$,
The Lippmann-Schwinger operator $T_{B_{R}}$ associated with the shifted body, written in momentum representation, is 
\begin{eqnarray*} &
T_{B_{R}}(q,k,\omega ')=\int \text{dxdy} e^{i (k\cdot x-q \cdot y)}T_B(x+{R}\hat{n},y+{R}\hat{n},\omega ')=\nn & \int \text{dxdy} e^{i (k \cdot x-q \cdot y)-i (k-q) \cdot {R}\hat{n}}T_B(x,y,\omega ')=\nn & e^{-i(k-q) \cdot {R}\hat{n} }T_B(q,k,\omega ')
\end{eqnarray*}
{We therefore write for the shifted position:}
\bea&
K(R)=\nn & \int \text{d}\omega\text{d$\omega $}'\text{  }d^3k d^3q\text{   }\frac{1}{k^2+\omega ^2}T_A(k,q,\omega )\frac{1}{q^2+\omega ^2}\times \nn & \frac{e^{-i(k-q)\cdot R \hat{n} }}{q^2+\omega '^2}T_B(q,k,\omega ')\frac{1}{k^2+\omega '^2}\left(\frac{\omega ^2\omega '^2}{|k||q|}+|k||q|\right)\eea
{We now rescale all momenta and frequencies appearing in the integral by:  }
$
\tilde{k}=R k.
$
\begin{eqnarray*} & K=\int d\tilde{\omega } d\tilde{\omega }'\text{  }d^3\tilde{k} d^3\tilde{q}\text{  }\frac{1}{R^2} \frac{1}{\tilde{k}^2+\omega ^2}T_A\left(\frac{\tilde{k}}{R},\frac{\tilde{q}}{R},\frac{\tilde{\omega }}{R}\right)\times \nn & \frac{1}{\tilde{q}^2+\tilde{\omega }^2}\frac{e^{-i\left(\tilde{k}-\tilde{q}\right)\cdot \hat{n} }}{\tilde{q}^2+\tilde{\omega }'^2}T_B\left(\frac{\tilde{q}}{R},\frac{\tilde{k}}{R},\frac{\tilde{\omega }'}{R}\right)\frac{1}{\tilde{k}^2+\tilde{\omega }'^2}\left(\frac{\tilde{\omega }^2\tilde{\omega }'^2}{\left|\tilde{k}\right|\left|\tilde{q}\right|}+\left|\tilde{k}\right|\left|\tilde{q}\right|\right)\end{eqnarray*}
\text{Note that for }$R\to \infty${ we have }
$
T_B\left(\frac{\tilde{q}}{R},\frac{\tilde{k}}{R},\frac{\tilde{\omega }'}{R}\right)\longrightarrow \left( \frac{\tilde{\omega }'}{R}\right)^2 \chi _B\left(\frac{\tilde{q}}{R},\frac{\tilde{k}}{R};\frac{\tilde{\omega }'}{R}\right)$
{ since in this limit }
$\chi \to 0$
{ and }$T_B$
{ can be approximated as in the dilute limit. } For concreteness, let us take 
\bea&
\chi_A(x,\omega )=\theta _A(x)\frac{\omega_{pA}^2}{\omega ^2+\omega_0^2} \nn & \theta _A(x)=\left\{
\begin{array}{c}
1 ~~~~~~~ x\in A \\
 0 ~~~\text{otherwise}
\end{array}
\right.
\eea
\text{We can now carry out the frequency integrals yielding:}
\bea&
K=\frac{\omega_{\text{pA}}{}^2\omega_{pB}^2 }{R^2} \int d^3\tilde{k} d^3\tilde{q}\theta _A\left(\frac{\tilde{k}}{R},\frac{\tilde{q}}{R}\right) \theta _B\left(\frac{\tilde{q}}{R},\frac{\tilde{k}}{R}\right)e^{-i\left(\tilde{k}-\tilde{q}\right)\cdot \hat{n} }\times \nn & \frac{\pi ^2 \left(2 k^2 q^2+2 k q (k+q) R \omega _0+(k+q)^2 R^2 \omega _0{}^2\right)}{4 k q (k+q)^2 \left(k+R \omega _0\right){}^2 \left(q+R \omega _0\right){}^2}
\eea
{For $R\rightarrow\infty$  we can use the approximation }
\bea
\theta _A\left(\frac{\tilde{k}}{R},\frac{\tilde{q}}{R}\right)= \int _A\text{dx} e^{i \frac{x}{R}\cdot \left(\tilde{k}-\tilde{q}\right)}\sim V_{A} \eea
where $V_{A}$ is the volume of body $A$. {We therefore have:}
\bea&
K\sim V_{A}V_{B}\frac{\pi^2\omega _{{pA}}^2\omega _{{pB}}^2}{R^2}\int  d^3k d^3q \times \nn & \frac{\left(2 k^2 q^2+2 k q (k+q) R \omega_0+(k+q)^2 R^2 \omega_0^2\right)}{4 k q (k+q)^2 (k+R \omega_0)^2 (q+R \omega_0)^2}e^{-i(k-q)\cdot \hat{n} }
\eea
{We can now carry out the 3d angular integrals in the standard way, writing: }$
q\cdot \hat{n}=\cos\theta |q|$,{ we get:}
\bea&
K=\omega _{{pA}}^2\omega _{{pB}}^2 V_{A}V_{B}\frac{4\pi ^4}{R^2}\int _0^{\infty } \text{dk} \text{dq}\times \nn &  \frac{(2 k^2 q^2+2 k q (k+q) R \omega_0+(k+q)^2 R^2 \omega_0^2)}{ (k+q)^2 (k+R \omega_0)^2 (q+R \omega_0)^2}\sin (k)\sin (q)
\eea
At this point, it is convenient to rescale back the momenta, writing: 
\bea&
K=\omega _{{pA}}^2\omega _{{pB}}^2 V_{A}V_{B}\frac{4\pi ^4}{R^2}\int _0^{\infty } \text{dk} \text{dq}\times \nn & \frac{ \left(2 k^2 q^2 +2 k q (k +q ) +(k +q )^2 \right) }{ (k+q)^2 (k+1)^2 (q+1)^2}\sin (k{  }R \omega_0)\sin (q R \omega_0)
\eea
{To analyze this integral we  do a couple of integrations by parts according to:}
\bea&
\iint_0^L\text{dxdy} F(x,y)\partial_x \partial_yG(x,y)=\left(F(x,y)G(x,y)|_{x=0}^L\right)|_{y=0}^L-\nn & \int _0^L\text{dy}\left(\partial _yF(x,y)\right)G(x,y)|_{x=0}^L-\nn & \int _0^L\text{dx}\left(\left(\partial _xF(x,y)\right)G(x,y)|_{y=0}^L\right)+ \nn & \int _0^L\text{dx}\int _0^L\text{dy} \left(\partial _x\partial _yF(x,y)\right) G(x,y)
\eea
\text{In our case we take: }
\bea
F=\frac{ \left(2 k^2 q^2 +2 k q (k\text{  }+q ) +(k +q )^2 \right) }{\text{  }(k+q)^2 (k+1)^2 (q+1)^2}
\eea
{and }
\bea
G=\frac{1}{(R \omega_0)^2} \cos (k\text{  }R \omega_0)\cos (q R \omega_0)
.\eea
Note that $F=0${ at }
$k\to \infty${ at }$q\to \infty${ as well as }$F(0,0)=1${ and }$G(0,0)=\frac{1}{(R \omega_0)^2}$
{Thus we have:}
\begin{eqnarray}&\label{Kint}
K=V_AV_B\frac{4\pi ^4\omega_{pA}^2\omega_{pB}^2}{\omega_0^2R^4}+\nn & V_AV_B\frac{4\pi ^4\omega_{pA}^2\omega_{pB}^2}{\omega_0^2R^4}\Big(\int _0^{\infty } \text{dk}\text{dq}(\partial_k \partial _qF)\cos (k R \omega_0)\cos (q R \omega_0) \nn &+2\int_0^{\infty }\text{dk}( \partial _kF(k,0))\cos (k R \omega_0)\Big)
\end{eqnarray}
The remaining integrals in \eqref{Kint} can straightforwardly be shown to decay as $R^{-2}$ (using more integrations by parts), showing that:
\begin{eqnarray}&
K=V_AV_B\frac{4\pi ^4\omega_{pA}^2\omega_{pB}^2}{\omega_0^2R^4}+O({1\over R^{6}}),\end{eqnarray}
establishing the asymptotics \eqref{R4 asymptotic}.

The complete analysis we check that the additional terms in \eqref{S2R lowest} give a sub-leading correction to \eqref{R4 asymptotic}. Indeed,
the term
\bea&
\text{Tr}  {\cal H}_0 (\delta {\cal G}_{\text{AUB}}- \delta{\cal G}_A-\delta {\cal G}_B )\nn & = \text{Tr}   {\cal H}_0\int \text{d}\omega g_0\left(T_{\text{AUB}}-T_A-T_B\right)g_0 
\eea
gives us a $R^{-6}$ decay. The analysis goes as follows.
Using cyclicity of the trace this expression is the same as: 
\bea&
\int _0^{\infty }\text{d}\omega \text{Tr} \left(T_{\text{AUB}}-T_A-T_B\right)g_0 {\cal H}_0g_0\eea
where we also used that $[g_{0},{\cal H}_{0}]=0$.
Using \eqref{corrpi free} for ${\cal H}_{0}$, we have
\bea&
\langle x|g_0(\omega ) {\cal H}_0g_0(\omega )|y\rangle = \nn & \int \frac{ d^3k}{(2\pi )^3}\text{   }|k| \frac{1}{\left(k^2+\omega ^2\right)^2}e^{i (x-y)\cdot k}=\nn & \frac{1}{4\pi ^2}\int _0^{\infty }\text{dk} \int _{-1}^1d X\frac{k^3}{\left(k^2+\omega ^2\right)^2}e^{i |x-y| X k}= \nn &  \frac{1}{2\pi ^2|x-y| } \int _0^{\infty }\text{dk}\text{   }\frac{k^2}{\left(k^2+\omega ^2\right)^2} \sin(|x-y| k)=\nn &\frac{1}{2\pi ^2\omega  |x-y| }\int _0^{\infty }d u \frac{u^2}{\left(u^2+1\right)^2} \sin(|x-y|  \omega  u)
\eea
In the dilute approximation:
\bea& 
\int _0^{\infty }\text{d}\omega \text{Tr} \left(T_{\text{AUB}}-T_A-T_B\right)g_0 {\cal H}_0g_0\sim\nn &  \int _0^{\infty }\text{d}\omega\int _{A\times B} \text{dx} \text{dy} \omega ^4\chi _A(i \omega )g_{0\text{AB}}(x,y)\chi _B(i \omega )\times \nn &  \langle x|g_0(\omega ) {\cal H}_0g_0(\omega )|y\rangle = \nn &  \int _{A\times B} \text{dx} \text{dy} \int _0^{\infty }\text{d}\omega \chi _A(i \omega )\chi _B(i \omega ) \times \nn & \int _0^{\infty }d u \frac{u^2}{\left(u^2+1\right)^2} \sin(|x-y| \omega  u)\frac{\omega ^3e^{-\omega |x-y| }}{(2\pi )^3|x-y|^2}=\nn & \int _{A\times B} \text{dx} \text{dy} \frac{1}{(2\pi )^3}\frac{1}{|x-y|^6} \int _0^{\infty }\text{d}\omega \chi _A\left(i \frac{\omega }{ |x-y| }\right)\chi _B\left(i \frac{\omega }{ |x-y| }\right)\times \nn &  \int _0^{\infty }d u \frac{u^2}{\left(u^2+1\right)^2} \sin( \omega  u)\omega ^3e^{-\omega  }\longrightarrow_{R\to \infty } \nn & \int _{A\times B} \text{dx} \text{dy} \frac{1}{(2\pi )^3}\frac{\chi _A(0)\chi _B(0)}{|x-y|^6} \int _0^{\infty }\text{d}\omega \times \nn &\int _0^{\infty }d u \frac{u^2}{\left(u^2+1\right)^2} \sin( \omega  u)\omega ^3e^{-\omega  }=\nn & {\omega _{{pA}}^2\omega _{{pB}}^2\over 5(2\pi )^3\omega _{{0}}^4}\int _{A\times B} \text{dx} \text{dy} \frac{1}{|x-y|^6}
\eea
giving us a contribution of $
\frac{1}{5(2\pi )^3R^6}\sim \frac{1}{1240 R^6}$, which is sub-leading to the $O(R^{-4})$ contribution \eqref{R4 asymptotic contribution}. 
%In a similar manner various terms are checked. The calculations are somewhat tedious and I will not report all of them here. 

\section{Discussion}
In this paper we continued the investigation initiated in \cite{klich2012prlentanglement} of the state of a field interacting with a dispersive medium within the Gaussian model. We showed that the state cannot be considered as thermal, but rather as a state where photons have a $k$ dependent effective temperature. We found that the field is described by a density matrix whose Von-Neumann
entropy diverges as described by eq. \eqref{Results entropy cutoff}.  

In addition to supplying details on some of the calculations carried out in \cite{klich2012prlentanglement} we present several new results: Namely, formulae for the distance dependent Renyi entropy , as well as the distance dependence of ${\cal S}_{2,R}$. We find that the decay in 3d is proportional to ${\cal S}_{2,R}\propto R^{-4}$.

This result is curious, in that it seems at odds with the scaling ${\cal S}_{R}\sim R^{-6}$ of the entropy for parallel plates per unit area found in  \cite{klich2012prlentanglement}. Indeed, using the asymptotic result \eqref{R4 asymptotic} we can approximate the distance dependent part of the entropy of two plates per unit area by a pair-summation as:
\bea&
{\cal S}_{2,R}\propto \int \text{dx} \text{dy} \frac{1}{\left(R^2+x^2+y^2\right)^2}= \nn &2\pi \int _0^{\infty }r\text{dr} \frac{1}{\left(R^2+r^2\right)^2}=\frac{\pi }{R^2}
\eea
giving us a much slower decay compared to that found in  \cite{klich2012prlentanglement}. It must be noted that the exactly solvable toy model considered in \cite{klich2012prlentanglement} is unusual in that it has a $\dot{\psi}\phi$ coupling rather than the $\dot{\phi}\psi$ coupling considered here (See the discussion in section II), but it is not clear if this difference is the source of the different scaling behavior. Also, the calculation in \cite{klich2012prlentanglement} was done for the full Von-Neumann entropy rather than ${\cal S}_{2,R}$ considered here. More work is needed to understand the difference between these results.
%The occupation number of modes behaves as $n_{k}\sim 1/\sqrt{k}$ per unit volume for "soft" photons. A similar situation may arise when considering the phonons in solid. 
%This effect is reminiscent of the infra red problem in quantum electrodynamics where,  infinite numbers of soft photons are generated in transition amplitudes (see, e.g. \cite{kibble1968coherent}).

We expect that much more insight into the mixed state of the electromagnetic field  may be gained using numerical means to study 
how other factors, such as geometries and vector properties, affect the Casimir entanglement entropy and entanglement spectrum. 

%\new Add refs to : balasubramanian, fradkin-moore, Gaussian channel?\old
{\bf Acknowledgments:} Financial support from NSF CAREER award No. DMR-0956053 and NSF grant No. NSF PHY11-25915 is gratefully acknowledged.

\bibliographystyle{apsrev}
\bibliography{/Users/iklich/dropbox/Work/KlichBib.bib}

\begin{thebibliography}{25}
\expandafter\ifx\csname natexlab\endcsname\relax\def\natexlab#1{#1}\fi
\expandafter\ifx\csname bibnamefont\endcsname\relax
  \def\bibnamefont#1{#1}\fi
\expandafter\ifx\csname bibfnamefont\endcsname\relax
  \def\bibfnamefont#1{#1}\fi
\expandafter\ifx\csname citenamefont\endcsname\relax
  \def\citenamefont#1{#1}\fi
\expandafter\ifx\csname url\endcsname\relax
  \def\url#1{\texttt{#1}}\fi
\expandafter\ifx\csname urlprefix\endcsname\relax\def\urlprefix{URL }\fi
\providecommand{\bibinfo}[2]{#2}
\providecommand{\eprint}[2][]{\url{#2}}

\bibitem[{\citenamefont{Goldstein et~al.}(2006)\citenamefont{Goldstein,
  Lebowitz, Tumulka, and Zangh{\'\i}}}]{goldstein2006canonical}
\bibinfo{author}{\bibfnamefont{S.}~\bibnamefont{Goldstein}},
  \bibinfo{author}{\bibfnamefont{J.}~\bibnamefont{Lebowitz}},
  \bibinfo{author}{\bibfnamefont{R.}~\bibnamefont{Tumulka}}, \bibnamefont{and}
  \bibinfo{author}{\bibfnamefont{N.}~\bibnamefont{Zangh{\'\i}}},
  \bibinfo{journal}{Physical review letters} \textbf{\bibinfo{volume}{96}},
  \bibinfo{pages}{50403} (\bibinfo{year}{2006}).

\bibitem[{\citenamefont{Popescu et~al.}(2006)\citenamefont{Popescu, Short, and
  Winter}}]{popescu2006entanglement}
\bibinfo{author}{\bibfnamefont{S.}~\bibnamefont{Popescu}},
  \bibinfo{author}{\bibfnamefont{A.}~\bibnamefont{Short}}, \bibnamefont{and}
  \bibinfo{author}{\bibfnamefont{A.}~\bibnamefont{Winter}},
  \bibinfo{journal}{Nature Physics} \textbf{\bibinfo{volume}{2}},
  \bibinfo{pages}{754} (\bibinfo{year}{2006}).

\bibitem[{\citenamefont{Klich}(2012)}]{klich2012prlentanglement}
\bibinfo{author}{\bibfnamefont{I.}~\bibnamefont{Klich}},
  \bibinfo{journal}{Phys. Rev. Lett.} \textbf{\bibinfo{volume}{109}},
  \bibinfo{pages}{061601} (\bibinfo{year}{2012}),
  \urlprefix\url{http://link.aps.org/doi/10.1103/PhysRevLett.109.061601}.

\bibitem[{\citenamefont{Lifshitz and Pitaevskii}(1984)}]{LP}
\bibinfo{author}{\bibfnamefont{E.~M.} \bibnamefont{Lifshitz}} \bibnamefont{and}
  \bibinfo{author}{\bibfnamefont{L.~P.} \bibnamefont{Pitaevskii}},
  \emph{\bibinfo{title}{Statistical Mechanics, Part 2}}
  (\bibinfo{publisher}{Pergamon}, \bibinfo{address}{Oxford},
  \bibinfo{year}{1984}).

\bibitem[{\citenamefont{Caldeira and Leggett}(1981)}]{caldeira1981influence}
\bibinfo{author}{\bibfnamefont{A.}~\bibnamefont{Caldeira}} \bibnamefont{and}
  \bibinfo{author}{\bibfnamefont{A.}~\bibnamefont{Leggett}},
  \bibinfo{journal}{Phys. Rev. Lett.} \textbf{\bibinfo{volume}{46}},
  \bibinfo{pages}{211} (\bibinfo{year}{1981}).

\bibitem[{\citenamefont{Le~Hur}(2008)}]{hur2008entanglement}
\bibinfo{author}{\bibfnamefont{K.}~\bibnamefont{Le~Hur}},
  \bibinfo{journal}{Annals of Physics} \textbf{\bibinfo{volume}{323}},
  \bibinfo{pages}{2208} (\bibinfo{year}{2008}).

\bibitem[{\citenamefont{Lambert et~al.}(2004)\citenamefont{Lambert, Emary, and
  Brandes}}]{lambert2004entanglement}
\bibinfo{author}{\bibfnamefont{N.}~\bibnamefont{Lambert}},
  \bibinfo{author}{\bibfnamefont{C.}~\bibnamefont{Emary}}, \bibnamefont{and}
  \bibinfo{author}{\bibfnamefont{T.}~\bibnamefont{Brandes}},
  \bibinfo{journal}{Physical review letters} \textbf{\bibinfo{volume}{92}},
  \bibinfo{pages}{73602} (\bibinfo{year}{2004}).

\bibitem[{\citenamefont{Marcovitch et~al.}(2009)\citenamefont{Marcovitch,
  Retzker, Plenio, and Reznik}}]{marcovitch2009critical}
\bibinfo{author}{\bibfnamefont{S.}~\bibnamefont{Marcovitch}},
  \bibinfo{author}{\bibfnamefont{A.}~\bibnamefont{Retzker}},
  \bibinfo{author}{\bibfnamefont{M.}~\bibnamefont{Plenio}}, \bibnamefont{and}
  \bibinfo{author}{\bibfnamefont{B.}~\bibnamefont{Reznik}},
  \bibinfo{journal}{Physical Review A} \textbf{\bibinfo{volume}{80}},
  \bibinfo{pages}{012325} (\bibinfo{year}{2009}).

\bibitem[{\citenamefont{Calabrese et~al.}(2009)\citenamefont{Calabrese, Cardy,
  and Tonni}}]{calabrese2009entanglement}
\bibinfo{author}{\bibfnamefont{P.}~\bibnamefont{Calabrese}},
  \bibinfo{author}{\bibfnamefont{J.}~\bibnamefont{Cardy}}, \bibnamefont{and}
  \bibinfo{author}{\bibfnamefont{E.}~\bibnamefont{Tonni}}, \bibinfo{journal}{J.
  of Stat. Phys.} \textbf{\bibinfo{volume}{2009}}, \bibinfo{pages}{P11001}
  (\bibinfo{year}{2009}).

\bibitem[{\citenamefont{Ju et~al.}(2012)\citenamefont{Ju, Kallin, Fendley,
  Hastings, and Melko}}]{ju2012entanglement}
\bibinfo{author}{\bibfnamefont{H.}~\bibnamefont{Ju}},
  \bibinfo{author}{\bibfnamefont{A.}~\bibnamefont{Kallin}},
  \bibinfo{author}{\bibfnamefont{P.}~\bibnamefont{Fendley}},
  \bibinfo{author}{\bibfnamefont{M.}~\bibnamefont{Hastings}}, \bibnamefont{and}
  \bibinfo{author}{\bibfnamefont{R.}~\bibnamefont{Melko}},
  \bibinfo{journal}{Physical Review B} \textbf{\bibinfo{volume}{85}},
  \bibinfo{pages}{165121} (\bibinfo{year}{2012}).

\bibitem[{\citenamefont{Hastings et~al.}(2010)\citenamefont{Hastings,
  Gonz{\'a}lez, Kallin, and Melko}}]{hastings2010measuring}
\bibinfo{author}{\bibfnamefont{M.}~\bibnamefont{Hastings}},
  \bibinfo{author}{\bibfnamefont{I.}~\bibnamefont{Gonz{\'a}lez}},
  \bibinfo{author}{\bibfnamefont{A.}~\bibnamefont{Kallin}}, \bibnamefont{and}
  \bibinfo{author}{\bibfnamefont{R.}~\bibnamefont{Melko}},
  \bibinfo{journal}{Physical review letters} \textbf{\bibinfo{volume}{104}},
  \bibinfo{pages}{157201} (\bibinfo{year}{2010}).

\bibitem[{\citenamefont{Song et~al.}(2011)\citenamefont{Song, Laflorencie,
  Rachel, and Le~Hur}}]{song2011entanglementHeisenberg}
\bibinfo{author}{\bibfnamefont{H.}~\bibnamefont{Song}},
  \bibinfo{author}{\bibfnamefont{N.}~\bibnamefont{Laflorencie}},
  \bibinfo{author}{\bibfnamefont{S.}~\bibnamefont{Rachel}}, \bibnamefont{and}
  \bibinfo{author}{\bibfnamefont{K.}~\bibnamefont{Le~Hur}},
  \bibinfo{journal}{Physical Review B} \textbf{\bibinfo{volume}{83}},
  \bibinfo{pages}{224410} (\bibinfo{year}{2011}).

\bibitem[{\citenamefont{Levin and Rytov}(1967)}]{levin1967theory}
\bibinfo{author}{\bibfnamefont{M.}~\bibnamefont{Levin}} \bibnamefont{and}
  \bibinfo{author}{\bibfnamefont{S.}~\bibnamefont{Rytov}},
  \bibinfo{journal}{Science, Moscow} \textbf{\bibinfo{volume}{6}}
  (\bibinfo{year}{1967}).

\bibitem[{\citenamefont{Botero and Reznik}(2003)}]{botero2003modewise}
\bibinfo{author}{\bibfnamefont{A.}~\bibnamefont{Botero}} \bibnamefont{and}
  \bibinfo{author}{\bibfnamefont{B.}~\bibnamefont{Reznik}},
  \bibinfo{journal}{Phys. Rev. A} \textbf{\bibinfo{volume}{67}},
  \bibinfo{pages}{052311} (\bibinfo{year}{2003}).

\bibitem[{\citenamefont{Plenio and Virmani}(2007)}]{plenio2007introduction}
\bibinfo{author}{\bibfnamefont{M.}~\bibnamefont{Plenio}} \bibnamefont{and}
  \bibinfo{author}{\bibfnamefont{S.}~\bibnamefont{Virmani}},
  \bibinfo{journal}{Quant. Inf. Comp.} \textbf{\bibinfo{volume}{7}},
  \bibinfo{pages}{1} (\bibinfo{year}{2007}).

\bibitem[{\citenamefont{Bezerra et~al.}(2004)\citenamefont{Bezerra,
  Klimchitskaya, Mostepanenko, and Romero}}]{bezerra2004violation}
\bibinfo{author}{\bibfnamefont{V.}~\bibnamefont{Bezerra}},
  \bibinfo{author}{\bibfnamefont{G.}~\bibnamefont{Klimchitskaya}},
  \bibinfo{author}{\bibfnamefont{V.}~\bibnamefont{Mostepanenko}},
  \bibnamefont{and} \bibinfo{author}{\bibfnamefont{C.}~\bibnamefont{Romero}},
  \bibinfo{journal}{Physical Review A} \textbf{\bibinfo{volume}{69}},
  \bibinfo{pages}{022119} (\bibinfo{year}{2004}).

\bibitem[{\citenamefont{Decca et~al.}(2005)\citenamefont{Decca, L{\'o}pez,
  Fischbach, Klimchitskaya, Krause, and Mostepanenko}}]{decca2005precise}
\bibinfo{author}{\bibfnamefont{R.}~\bibnamefont{Decca}},
  \bibinfo{author}{\bibfnamefont{D.}~\bibnamefont{L{\'o}pez}},
  \bibinfo{author}{\bibfnamefont{E.}~\bibnamefont{Fischbach}},
  \bibinfo{author}{\bibfnamefont{G.}~\bibnamefont{Klimchitskaya}},
  \bibinfo{author}{\bibfnamefont{D.}~\bibnamefont{Krause}}, \bibnamefont{and}
  \bibinfo{author}{\bibfnamefont{V.}~\bibnamefont{Mostepanenko}},
  \bibinfo{journal}{Annals of Physics} \textbf{\bibinfo{volume}{318}},
  \bibinfo{pages}{37} (\bibinfo{year}{2005}).

\bibitem[{\citenamefont{Brevik et~al.}(2005)\citenamefont{Brevik, Aarseth,
  H{\o}ye, and Milton}}]{brevik2005temperature}
\bibinfo{author}{\bibfnamefont{I.}~\bibnamefont{Brevik}},
  \bibinfo{author}{\bibfnamefont{J.}~\bibnamefont{Aarseth}},
  \bibinfo{author}{\bibfnamefont{J.}~\bibnamefont{H{\o}ye}}, \bibnamefont{and}
  \bibinfo{author}{\bibfnamefont{K.}~\bibnamefont{Milton}},
  \bibinfo{journal}{Physical Review E} \textbf{\bibinfo{volume}{71}},
  \bibinfo{pages}{056101} (\bibinfo{year}{2005}).

\bibitem[{\citenamefont{Sushkov et~al.}(2011)\citenamefont{Sushkov, Kim,
  Dalvit, and Lamoreaux}}]{Sushkov:2011fk}
\bibinfo{author}{\bibfnamefont{A.~O.} \bibnamefont{Sushkov}},
  \bibinfo{author}{\bibfnamefont{W.~J.} \bibnamefont{Kim}},
  \bibinfo{author}{\bibfnamefont{D.~A.~R.} \bibnamefont{Dalvit}},
  \bibnamefont{and} \bibinfo{author}{\bibfnamefont{S.~K.}
  \bibnamefont{Lamoreaux}}, \bibinfo{journal}{Nat. Phys.}
  \textbf{\bibinfo{volume}{7}}, \bibinfo{pages}{230} (\bibinfo{year}{2011}).

\bibitem[{\citenamefont{Klingshirn}(2005)}]{klingshirn2005semiconductor}
\bibinfo{author}{\bibfnamefont{C.}~\bibnamefont{Klingshirn}},
  \emph{\bibinfo{title}{Semiconductor optics}}, vol. \bibinfo{volume}{1439}
  (\bibinfo{publisher}{Springer Verlag}, \bibinfo{year}{2005}).

\bibitem[{\citenamefont{Araki and Lieb}(1970)}]{araki1970entropy}
\bibinfo{author}{\bibfnamefont{H.}~\bibnamefont{Araki}} \bibnamefont{and}
  \bibinfo{author}{\bibfnamefont{E.}~\bibnamefont{Lieb}},
  \bibinfo{journal}{Comm. Math. Phys.} \textbf{\bibinfo{volume}{18}},
  \bibinfo{pages}{160} (\bibinfo{year}{1970}), ISSN \bibinfo{issn}{0010-3616}.

\bibitem[{\citenamefont{Geyer et~al.}(2005)\citenamefont{Geyer, Klimchitskaya,
  and Mostepanenko}}]{Geyer:2005ap}
\bibinfo{author}{\bibfnamefont{B.}~\bibnamefont{Geyer}},
  \bibinfo{author}{\bibfnamefont{G.}~\bibnamefont{Klimchitskaya}},
  \bibnamefont{and}
  \bibinfo{author}{\bibfnamefont{V.}~\bibnamefont{Mostepanenko}},
  \bibinfo{journal}{Phys.Rev.} \textbf{\bibinfo{volume}{D72}},
  \bibinfo{pages}{085009} (\bibinfo{year}{2005}).

\bibitem[{\citenamefont{Bordag and Pirozhenko}(2010)}]{bordag2010casimir}
\bibinfo{author}{\bibfnamefont{M.}~\bibnamefont{Bordag}} \bibnamefont{and}
  \bibinfo{author}{\bibfnamefont{I.}~\bibnamefont{Pirozhenko}},
  \bibinfo{journal}{Phys. Rev. D.} \textbf{\bibinfo{volume}{82}},
  \bibinfo{pages}{125016} (\bibinfo{year}{2010}).

\bibitem[{\citenamefont{Feinberg et~al.}(2001)\citenamefont{Feinberg, Mann, and
  Revzen}}]{feinberg2001casimir}
\bibinfo{author}{\bibfnamefont{J.}~\bibnamefont{Feinberg}},
  \bibinfo{author}{\bibfnamefont{A.}~\bibnamefont{Mann}}, \bibnamefont{and}
  \bibinfo{author}{\bibfnamefont{M.}~\bibnamefont{Revzen}},
  \bibinfo{journal}{Annals of Physics} \textbf{\bibinfo{volume}{288}},
  \bibinfo{pages}{103} (\bibinfo{year}{2001}).

\bibitem[{\citenamefont{Kenneth and Klich}(2008)}]{kenneth2008casimir}
\bibinfo{author}{\bibfnamefont{O.}~\bibnamefont{Kenneth}} \bibnamefont{and}
  \bibinfo{author}{\bibfnamefont{I.}~\bibnamefont{Klich}},
  \bibinfo{journal}{Physical Review B} \textbf{\bibinfo{volume}{78}},
  \bibinfo{pages}{14103} (\bibinfo{year}{2008}).

\end{thebibliography}
\end{document}